\documentclass[preprint2]{aastex62}
\usepackage{amssymb}
\usepackage{amsmath}

\received{Aug 6, 2019}
\revised{Jan 14, 2019; Apr 19, 2019; May 15, 2019}
\accepted{May 16, 2019}
\submitjournal{The Astrophysical Journal}

\begin{document}
\title{The Effect of 3D Transport-Induced Disequilibrium Carbon Chemistry on the Atmospheric Structure and Phase Curves and Emission Spectra of Hot Jupiter HD 189733b}
\author{Maria E. Steinrueck}
\affil{Lunar and Planetary Laboratory, University of Arizona, Tucson, AZ, 85721, USA}
\email{msteinru@lpl.arizona.edu}

\author{Vivien Parmentier}
\affil{Department of Physics, University of Oxford, Oxford OX1 3PU, UK}

\author{Adam P. Showman}
\affil{Lunar and Planetary Laboratory, University of Arizona, Tucson, AZ, 85721, USA}

\author{Joshua D. Lothringer}
\affil{Lunar and Planetary Laboratory, University of Arizona, Tucson, AZ, 85721, USA}

\author{Roxana E. Lupu}
\affil{BAER Institute/NASA Ames Research Center, Moffett Field, CA 94043, USA}

\begin{abstract}
On hot Jupiter exoplanets, strong horizontal and vertical winds should homogenize the abundances of the important absorbers CH$_4$ and CO much faster than chemical reactions restore chemical equilibrium. This effect, typically neglected in general circulation models (GCMs), has been suggested as explanation for discrepancies between observed infrared lightcurves and those predicted by GCMs: On the nightsides of several hot Jupiters, GCMs predict outgoing fluxes that are too large, especially in the Spitzer 4.5 $\mu$m band.  We modified the SPARC/MITgcm to include disequilibrium abundances of CH$_4$, CO and H$_2$O by assuming that the CH$_4$/CO ratio is constant throughout the simulation domain. We ran simulations of hot Jupiter HD 189733b with 8 CH$_4$/CO ratios. In the more likely CO-dominated regime, we find temperature changes $\geq$50-100 K compared to the equilibrium chemistry case across large regions. This effect is large enough to affect predicted emission spectra and should thus be included in GCMs of hot Jupiters with equilibrium temperatures between 600K and 1300K. We find that spectra in regions with strong methane absorption, including the Spitzer 3.6 and 8 $\mu$m bands, are strongly impacted by disequilibrium abundances. We expect chemical quenching to result in much larger nightside fluxes in the 3.6 $\mu$m band, in stark contrast to observations. Meanwhile, we find almost no effect on predicted observations in the 4.5 $\mu$m band, as the opacity changes due to CO and H$_2$O offset each other. We thus conclude that disequilibrium carbon chemistry cannot explain the observed low nightside fluxes in the 4.5 $\mu$m band.
\end{abstract}


\section{Introduction}
Close-in extrasolar giant planets, known as hot Jupiters, are the best characterized exoplanets to date. Due to their proximity to their host stars, they are expected to be tidally locked. This creates strong temperature contrasts between the permanent day side and the night side. For the temperature ranges of typical hot Jupiters, assuming chemical equilibrium, these temperature contrasts translate to large horizontal gradients in the abundance of methane (CH$_4$) and carbon monoxide (CO), two important infrared absorbers in the atmospheres of hot Jupiters. Carbon is preferentially found in CH$_4$ at high pressures and low temperatures, while at low pressures and high temperatures, CO is the dominant carbon-bearing species. Most models of hot Jupiters, especially three-dimensional general circulation models (GCMs) allowing for realistic representation of opacities and the effect that chemical composition exert on them, assume equilibrium chemistry as a basis for calculating opacities. However, the stark day-night temperature contrast also drives a vigorous atmospheric circulation, with a strong eastward equatorial jet advecting thermal energy from the day side to the night side and strong vertical mixing \citep[e.g.][]{ShowmanGuillot2002,ShowmanPolvani2011,Dobbs-DixonLin2008,HengEtAl2011b,ThrastarsonCho2011,RauscherMenou2012,PernaEtAl2012,Dobbs-DixonAgol2013,ParmentierEtAl2013,MayneEtAl2014,KatariaEtAl2016,MendoncaEtAl2016}. 
In addition, at low pressures ($\lesssim 1$ bar), the chemical time scale at which the interconversion between CH$_4$ and CO acts, becomes very long. Therefore, an air parcel is advected much faster than its CH$_4$ and CO abundances can adapt to the local equilibrium values. This process, called quenching, is expected to vertically and horizontally homogenize the abundances of CH$_4$ and CO (as well as of many other species, including N$_2$ and NH$_3$) in the near-infrared photospheres (located roughly between 1 bar and 1 mbar) of hot Jupiters \citep{CooperShowman2006,AgundezEtAl2012,AgundezEtAl2014,DrummondEtAl2018,DrummondEtAl2018HD189733b}.

These disequilibrium abundances can have a significant effect on the opacities and thus the radiative transfer. Including this effect in general circulation models (GCMs) could potentially impact the thermal structure and atmospheric circulation, as well as the predicted spectra and phase curves. In fact, this has been proposed as a solution for the observed discrepancy between phase curves predicted by state-of-the-art GCMs assuming equilibrium chemistry and observations with the Spitzer Space Telescope \citep{KnutsonEtAl2012}. GCMs over-predict the night side fluxes in the Spitzer 4.5 $\mu$m band for the hot Jupiters HD 189733b \citep{KnutsonEtAl2012}, HD 209458b \citep{ZellemEtAl2014} and WASP-43b \citep{StevensonEtAl2017}. Within this wavelength bandpass, CO has a strong absorption band. For relatively cool hot Jupiters, like HD 189733b, transport-induced disequilibrium chemistry is expected to enhance the CO abundance on the night side compared to the equilibrium chemistry value \citep{CooperShowman2006,AgundezEtAl2014}. \citet{KnutsonEtAl2012} argued that this would lead to increased opacity in the 4.5 $\mu$m band, such that the outgoing radiation in that band would probe higher, cooler regions of the atmosphere, decreasing the flux emitted in this band.
In addition to changing the pressure level from which the outgoing radiation is emitted, the altered opacities can also affect the thermal structure and the atmospheric circulation in a GCM, an effect not considered in the argumentation of \citet{KnutsonEtAl2012}. So far, this effect has only been taken into account by \citet{DrummondEtAl2018,DrummondEtAl2018HD189733b}.

The goal of this study is to better quantify how the combination of these two effects of disequilibrium carbon chemistry (the change in the level from which radiation escapes to space and the change in thermal structure) affects the thermal emission spectra predicted from GCMs. We assume that the abundance ratio of CH$_4$ to CO is constant throughout the entire simulation domain and treat the CH$_4$/CO abundance ratio as free parameter. This simple approach allows us to explore a broader parameter range than \citet{DrummondEtAl2018,DrummondEtAl2018HD189733b} and to focus on the radiative effects. Our approach is justified by the findings of previous studies: Coupling a simple chemical relaxation scheme to a GCM of HD 209458b, both \citet{CooperShowman2006} and \citet{DrummondEtAl2018} found that the CO and CH$_4$ abundances are homogenized everywhere above the $\sim 3$ bar level. In a later study (published while this paper was in the peer-review process), \citet{DrummondEtAl2018HD189733b} find that the same is true for HD 189733b.
 \citet{AgundezEtAl2014} instead used a full kinetical network in a pseudo-2D framework that was able to capture vertical and horizontal transport, and included photochemistry on the day side. They found that quenching homogenized the CO and CH$_4$ abundances at pressures between $\sim1$ and $\sim10^{-4}$ bars on HD 189733b. (At lower pressures, photochemical processes destroy CH$_4$ on the day side.) While these studies disagree on the relative importance of vertical versus horizontal quenching, all of them conclude that the abundances of CO and CH$_4$ should be homogeneous in the near-infrared photosphere, justifying our assumption.

However, coupling a chemical relaxation scheme to their GCM, \citet{MendoncaEtAl2018_chemistry} found that on WASP-43b, the CH$_4$ abundances were only homogenized horizontally but not vertically. A similar behavior is seen in the HD 209458b case of \citet{AgundezEtAl2014} (though only for their nominal eddy diffusion profile). While there are several differences in the models and planet parameters used that could contribute to this different outcome, the perhaps most important factors are the hotter day side and the weak vertical mixing in both of these models. 
If horizontal transport dominates over vertical transport in setting the disequilibrium abundances, as these two papers find, whether abundances are homogenized only horizontally or vertically and horizontally depends on whether vertical quenching happens in the hottest region of the day side. Abundances thus are homogenized vertically and horizontally if in the hottest regions of the day side the vertical mixing time scale is shorter than the chemical time scale. If vertical mixing is very weak or the day side is too hot (leading to a shorter chemical time scale), this condition is not fulfilled and abundances are only homogenized horizontally but not vertically. In the case of \citet{AgundezEtAl2014}, the hotter day side compared to \citet{DrummondEtAl2018} at low pressures is largely due to their assumption of a thermal inversion on the day side of HD 209458b. \citet{MendoncaEtAl2018_chemistry} look at WASP-43b, for which \citet{KatariaEtAl2015} also found a larger day-night contrast and hotter day side compared to HD 209458b due to its shorter orbital period and larger gravity. Our study looks at HD 189733b, which has a significantly lower zero-albedo equilibrium temperature than HD 209458b and WASP-43b. Since the chemical time scale dramatically increases with decreasing temperature, it is likely that abundances are homogenized horizontally and vertically on this planet, and both published studies looking at HD 189733b confirm this \citep{AgundezEtAl2014,DrummondEtAl2018HD189733b}.

 With the focus of these previous studies being on chemistry, most of them did not self-consistently calculate the radiative transfer, and thus were not able to quantify the effect of the changed opacities on the temperature structure in their model. In their GCM, \citet{CooperShowman2006} used a Newtonian cooling scheme, in which the temperature relaxes towards a prescribed temperature profile at each point in the atmosphere. Like typical kinetical networks, \citet{AgundezEtAl2014} used a prescribed background pressure-temperature profile (in this case derived from a GCM assuming equlibrium chemistry). \citet{MendoncaEtAl2018_chemistry} use a double-grey radiative transfer scheme. Only \citet{DrummondEtAl2018} and \citet{DrummondEtAl2018HD189733b}, whose GCM uses state-of-the-art radiative transfer with correlated k-coefficients, included the effect of the changed opacities on the temperature structure in a GCM. Our approach complements previous studies by focusing on the radiative transfer rather than on chemistry. The photospheric disequilibrium CH$_4$ and CO abundances found in coupled chemistry-circulation models such as \citet{DrummondEtAl2018} strongly depend on the temperature profile in transition region between equilibrium chemistry and disequilibrium chemistry ($\sim 1$ to 10 bars)\citep{MosesEtAl2011,VenotEtAl2014}. However, the temperature profile in this region depends on the initial temperature profile used in the GCM \citep{AmundsenEtAl2016}, assumptions about the interior heat flux \citep[e.g.,][]{GuillotShowman2002,BurrowsEtAl2003,FortneyEtAl2008} 
 and dissipation in deep layers \citep[e.g.,][]{GuillotShowman2002,TremblinEtAl2017, KomacekYoudin2017} . Furthermore, typical GCM simulations only resolve dynamic mixing through the large-scale circulation. If unresolved sub-grid-scale turbulence is relevant near the quench level \citep[e.g., ][]{Menou2019}, it might further alter the quenched abundances. Uncertainties in reaction rates can also affect the resulting abundances \citep[e.g., ][]{VisscherMoses2011}. Overall, although it is well established that the CH4/CO ratio should be reasonably homogenized throughout the photosphere, many unconstrained factors will determine the actual value of this ratio. Studying how the thermal response and the resulting emission spectra and phase curves depends on the CH$_4$/CO ratio, as is our goal, thus adds to the overall understanding of the effects of disequilibrium chemistry.

\begin{figure*}
\plotone{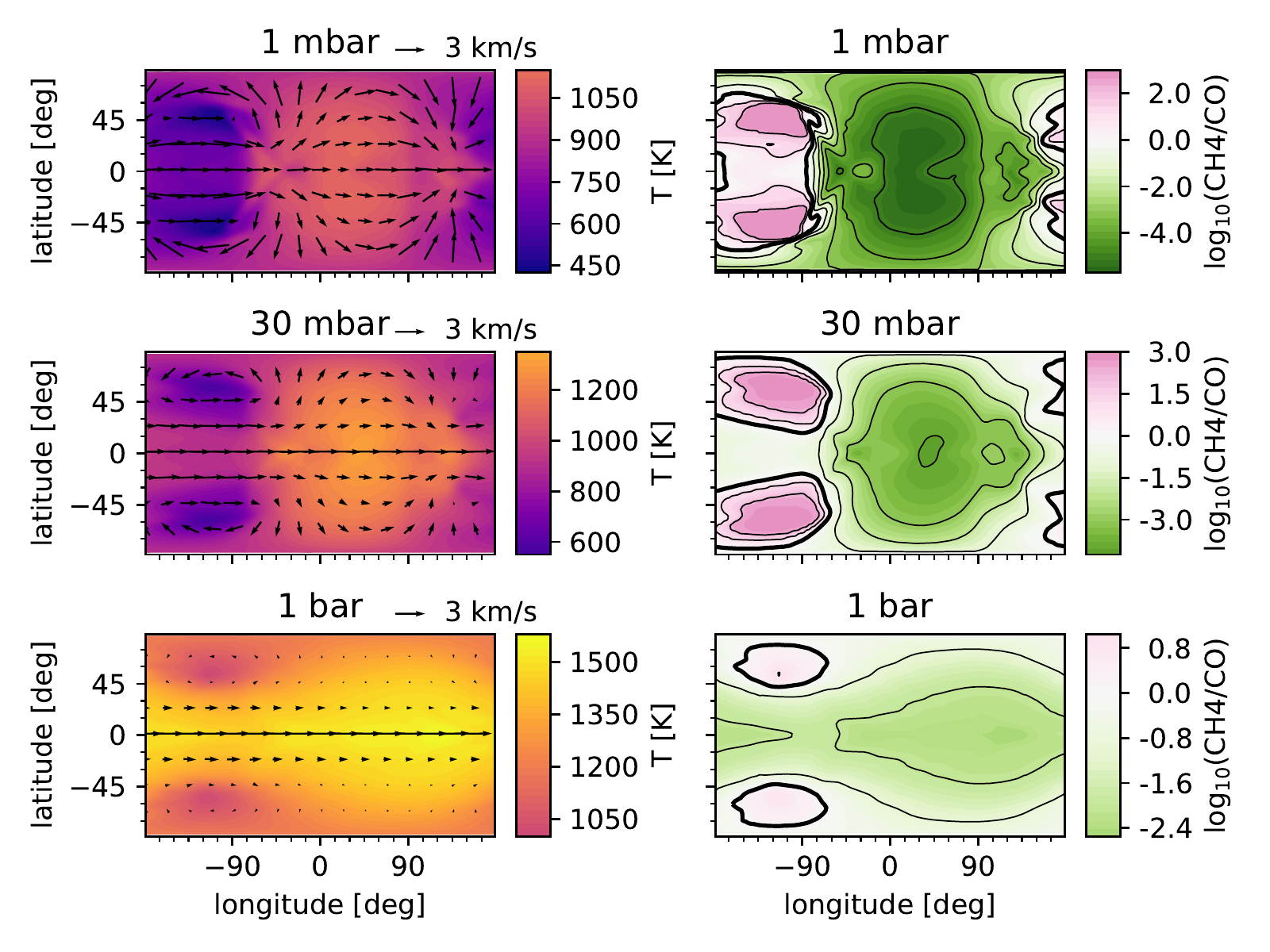}
\caption{Left: horizontal temperature maps of the reference simulation (equilibrium chemistry) at pressures of 1 mbar, 30 mbars and 1 bar, respectively. The substellar point (0 $^\circ$ longitude, 0 $^\circ$ latitude) is at the center of each panel. The arrows represent the velocities of the horizontal component of the wind. Right: horizontal maps of the abundance ratio of methane to carbon monoxide assuming chemical equilibrium. Methane dominated regions are magenta while carbon-monoxide dominated regions are green. The contours are evenly spaced in log space and mark log$_{10}$(CH$_4$/CO)=(-4,-3,-2,-1,0,1, 2), respectively. The zero contour (corresponding to CH$_4$/CO=1) is thickened. \label{fig:temp_composition}}
 
\end{figure*}

We choose to focus on HD 189733b. Due to its cooler temperature compared to HD 209458b and WASP-43b, disequilibrium effects are expected to be stronger on this planet. The latter two planets are hot enough that in simulations assuming equilibrium chemistry, CO is the dominant carbon species at all longitudes and latitudes \citep{ShowmanEtAl2009,DrummondEtAl2018}.
 In contrast, on the cooler HD 189733b, in chemical equilibrium one expects that the day side is dominated by CO, while the night side is dominated by CH$_4$ (see Figure \ref{fig:temp_composition}).  HD 189733b thus occupies an interesting point in the parameter space: Regardless of the quenched CH$_4$ and CO abundances, including disequilibrium chemistry will change which of the two species is dominant on about half of the planet. The radiative effects of disequilibrium chemistry are thus expected to play a larger role than on hotter planets. In addition, unlike for HD 209458b, for HD 189733b there exist phase curve observations in multiple Spitzer bandpasses \citep{KnutsonEtAl2007,KnutsonEtAl2009_HD189,AgolEtAl2010,KnutsonEtAl2012}, providing stronger observational constraints.

\section{Methods}
We use the SPARC/MITgcm model \citep{ShowmanEtAl2009} to perform simulations of hot Jupiter HD 189733b. This model couples the two-stream, non-gray radiative transfer code of \cite{MarleyMcKay1999}
to the MITgcm general circulation model of \cite{AdcroftEtAl2004} and has previously been applied to a wide range of hot Jupiters and other exoplanets 
\citep[e.g.,][]{ShowmanEtAl2009,ShowmanEtAl2013,ShowmanEtAl2015,ParmentierEtAl2013,ParmentierEtAl2018,LewisEtAl2010,LewisEtAl2014,KatariaEtAl2013,KatariaEtAl2014,KatariaEtAl2015}. 




\begin{deluxetable}{lrl}




\tablecaption{Model parameters\label{tab:modelparameters}}


\tablehead{\colhead{Parameter} & \colhead{Value} & \colhead{Units} \\ 
\colhead{} & \colhead{} & \colhead{} } 

\startdata
Radius & $7.9559\cdot 10^7$ & m \\
Gravity & 22.86 & m s$^{-2}$ \\
Rotation period & 2.22  & d \\
Semimajor axis & 0.03142 & AU \\
Specific heat capacity & $1.3\cdot 10^4$ & J kg$^{-1}$ K$^{-1}$ \\
Specific gas constant & 3714 & J kg$^{-1}$ K$^{-1}$ \\
Interior flux & 0 & W m$^{-2}$ \\
Horizontal resolution & C32\tablenotemark{a} &  \\
Vertical resolution & 53 & layers \\
Lower pressure boundary & $2\cdot 10^{-6}$ & bar \\
Upper pressure boundary & 200 & bar \\
Hydrodynamic time step & 25 & s \\
Radiative time step & 50 & s \\
\enddata

\tablenotetext{a}{equivalent to a resolution of 128x64 on a longitude-latitude grid}



\end{deluxetable}

\subsection{Dynamics}
Using the MITgcm \citep{AdcroftEtAl2004}, we solve the three-dimensional global primitive equations on a cubed sphere grid. The primitive equations are valid for stably stratified shallow atmospheres. Horizontal noise is smoothed with a fourth-order Shapiro filter \citep{Shapiro1970}.

The key model parameters are summarized in Table \ref{tab:modelparameters}. We assume a gravity of $g=22.86$ m s$^{-2}$, a planetary radius of $7.9559\cdot 10^7$ m 
and a rotation period equal to the orbital period of 2.22 days, as the planet is assumed to be in synchronous rotation. We use a value of $c_p=1.3\cdot 10^4$ J kg$^{-1}$ K$^{-1}$ for the specific heat capacity and $\kappa=R/c_p=2/7$, where $R$ is the specific gas constant. These values are appropriate for hydrogen dominated atmospheres.  Our simulations use a horizontal resolution of C32, which is roughly equivalent to a resolution of 128x64 on a longitude-latitude grid. The pressure spans from 200 bars to 2 $\mu$bar and is resolved by 53 vertical levels. We use a timestep of 25 s, run the simulations for 1,000 days and then average over the last 100 days. Convergence tests in previous studies using the SPARC/MITgcm have shown that simulations have sufficiently converged at that point. Specifically, the emitted infrared flux changes by less than a few percent between integration times of $\approx1,000$ days and $\approx4,000$ days. \citep[Fig. 12]{ShowmanEtAl2009}.

\subsection{Radiative transfer}
\label{subsec:radtran}
The radiative transfer code in our model is based on the plane-parallel code of \cite{MarleyMcKay1999}. This code was originally developed to study Titan's atmosphere \citep{McKayEtAl1989} and has since been applied to Uranus \citep{MarleyMcKay1999}
, brown dwarfs \citep{BurrowsEtAl1997,MarleyEtAl1996,MarleyEtAl2002} and hot Jupiters \citep{FortneyEtAl2005,FortneyEtAl2008,ShowmanEtAl2009}.
The code uses the delta-discrete ordinate method \citep{ToonEtAl1989} for the incident stellar flux, while the thermal flux is calculated using the two-stream  source function method \citep[also of][]{ToonEtAl1989}. Molecular opacities are treated using the correlated k-method \citep[e.g.][]{GoodyYung1989}: In each frequency bin, the opacity information from line-by-line calculations using as many as 10,000 to 100,000 frequency bins is turned into a cumulative distribution of opacities, which is then described by 8 k-coefficients. We use 11 frequency bins spanning from 0.26 $\mu$m to 325 $\mu$m \citep[see][]{KatariaEtAl2013}. The correlated-k method is the most sophisticated and accurate treatment of opacities in GCMs of exoplanets to date and is used both in the SPARC/MITgcm and in the adaptation for hot Jupiter of the UK MetOffice model \citep{AmundsenEtAl2016}.

In order to combine the opacities from different species within the atmosphere, we employ the approach of pre-mixed opacities, meaning that the k-coefficients of the mixture are derived from line-by-line opacities that are calculated for a mixture with specified chemical abundances that are specified functions of temperature and pressure (typically derived from equilibrium chemistry). This approach is fast and accurate. We use the equilibrium chemistry abundances for solar metallicity of \cite{LoddersFegley2002} and \cite{VisscherEtAl2006} with the modifications for CH$_4$, CO and H$_2$O described in section \ref{subsec:methods_disequilibriumchemistry} and the opacities of \cite{FreedmanEtAl2008} including the updates of \cite{FreedmanEtAl2014}. Note that the updated opacities result in changes in the thermal structure and the resulting equilibrium chemistry phase curve compared to \citet{ShowmanEtAl2009} and \citet{KnutsonEtAl2012} (see Section \ref{subsec:discussion_otherfactors}).
The disadvantage of this method is that it leaves little flexibility for varying abundances due to disequilibrium chemistry. However, alternative methods combining the k-coefficients of individual species on the fly come at a much higher computational cost and are thus impractical when coupled to GCMs \citep[as discussed in][]{AmundsenEtAl2017}. To explore the effect of disequilibrium abundances, we thus take advantage of the fact that due to strong horizontal and vertical mixing, the disequilibrium abundances of CH$_4$ and CO are expected to be close to constant throughout the photosphere, as further detailed in section \ref{subsec:methods_disequilibriumchemistry}.

To obtain spectra and phase curves, we postprocess our simulation output using the same plane-parallel two-stream radiative transfer code as used in our full 3D simulations, but adopting 196 wavelength bins instead of 11 to yield improved spectral resolution.  Given the time-averaged temperature structure from the GCM at a particular time, we calculate the outgoing flux  in the line of sight to the observer for each atmospheric column. The fluxes are then combined into a weighted average across the disk to give the total flux received at a particular observing vantage point and at a particular time during the orbit.  We do this calculation at many orbital phases (throughout which the Earth-facing hemisphere shifts in longitude) to assemble full orbit phase curves at wavelengths of interest. This method is similar to the method described in \cite{FortneyEtAl2006}, \citet{ShowmanEtAl2009} and \citet{ParmentierEtAl2016} and naturally takes into account limb darkening of the planet. We use a NextGen spectrum \citep{HauschildtEtAl1999}
for HD 189733, a stellar radius of 0.805 $R_\odot$ \citep{BoyajianEtAl2015} and a planet-to-star radius ratio  $R_p/R_\ast=0.0145$.

\subsection{Disequilibrium chemistry}
\label{subsec:methods_disequilibriumchemistry}
\begin{figure}
\epsscale{1.2}
\plotone{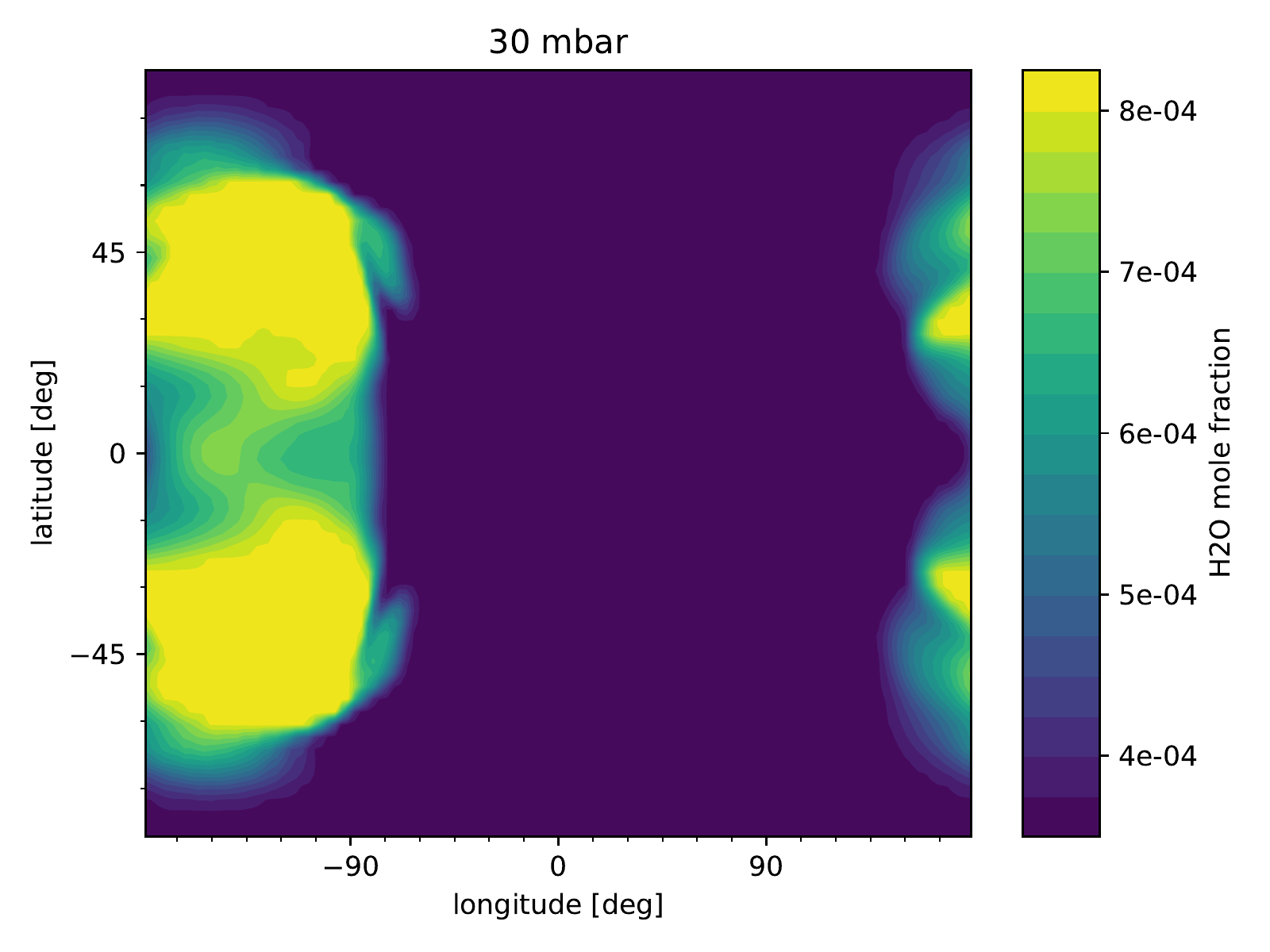}
\caption{Water abundance at the 30 mbar level in the equilibrium chemistry simulation. The substellar point is at the center of the panel. In disequilibrium chemistry, in contrast, the abundances are expected to be homogenized between the day- and night side of the planet.} \label{fig:h2oabundance}
\end{figure}

\begin{figure}
\epsscale{1.2}
\plotone{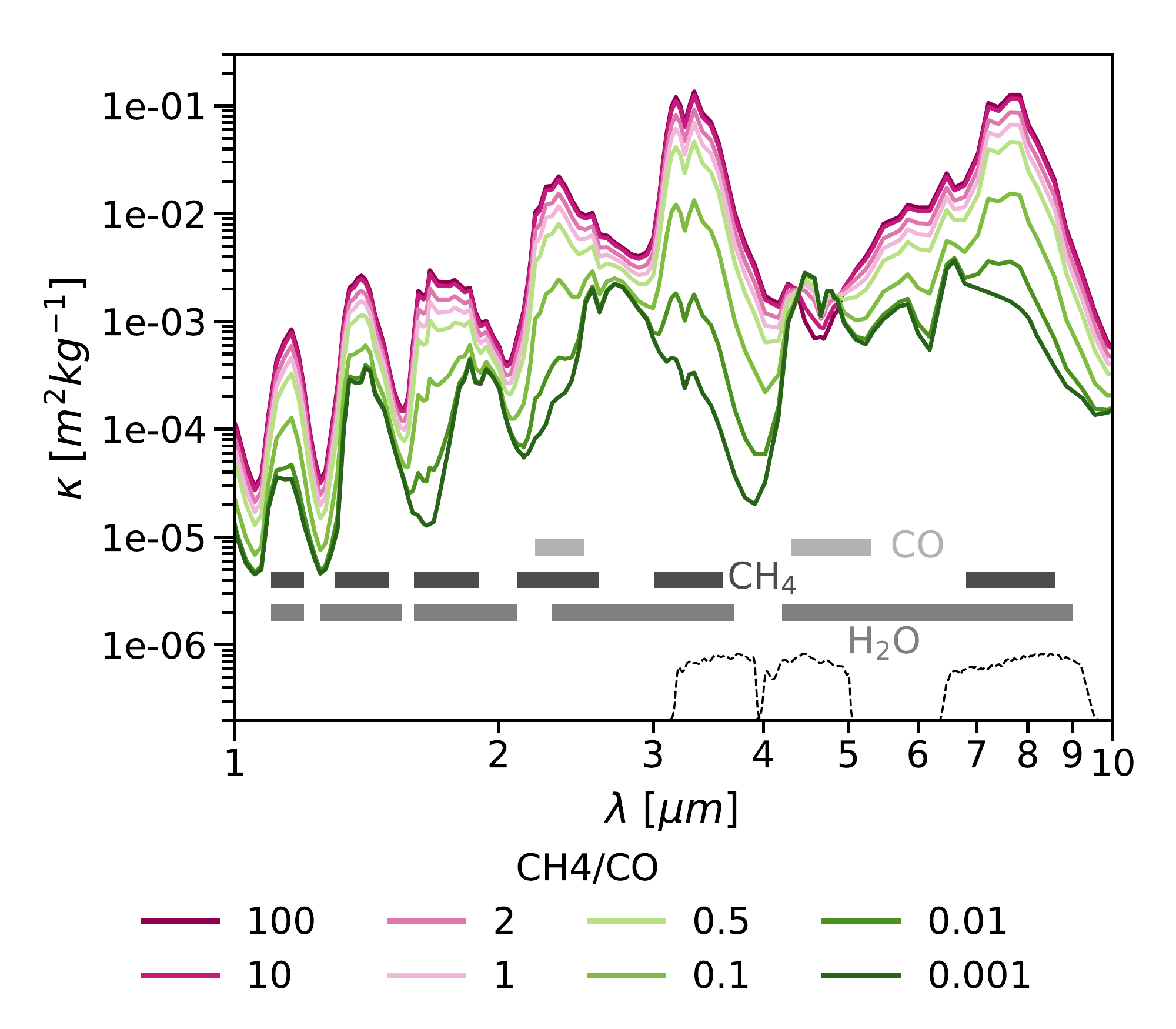}
\caption{Opacities for different CH$_4$/CO ratios at a temperature of 1000 K and a pressure of 30 mbar. Opacities for CH$_4$ dominated ratios are plotted in magenta and CO dominated ratios are plotted in green. The locations of imporant absorption bands are shown as grey bars towards the bottom of the figure. The dashed black lines indicate the filter sensitivity profiles for the Spitzer 3.6 $\mu$m, 4.5 $\mu$m and 8 $\mu$m bands. \label{fig:opacities}}
\end{figure}

All previous studies of hot Jupiters using the SPARC/MITgcm utilized k-coefficient tables calculated assuming equilibrium chemistry. To explore the effect of transport-induced disequilibrium carbon chemistry, in this study we instead assume that the CH$_4$ to CO ratio is quenched to a constant value throughout the entire atmosphere. This assumption is a good approximation at pressures lower than roughly 1 to 10 bars, as shown by \citet{CooperShowman2006}, \citet{AgundezEtAl2014} and  \citet{DrummondEtAl2018,DrummondEtAl2018HD189733b}.
At higher pressures, the chemical timescale becomes shorter than the mixing time scale and the approximation of a constant quenched value breaks down. Since this study focuses on the effect on the outgoing radiation in the near infrared, which probes pressures beween 1 mbar and 1 bar (see Fig. 9 in \citet{ShowmanEtAl2009} and Fig. 12 and 14 in \citet{MosesEtAl2011}, we believe this approximation is justified. 
The deep regions of the atmosphere (deeper than 10 bars) also have extremely small net fluxes compared to the fluxes in the observable atmosphere (where the incoming starlight is absorbed and radiation is escaping to space), and so the dynamics in the observable atmosphere are not strongly sensitive to modest changes in the opacities and fluxes in the deep ($p>$10 bar) atmosphere. In addition, because of the low net fluxes, the error in opacity we make by extending constant CH$_4$/CO ratios to the deep atmosphere is expected to have a relatively small effect on the temperature structure. We expect the error in the temperature profile in the deep atmosphere to be smaller than, or at most comparable to, the uncertainty due to the unknown internal heat flux of hot Jupiters and certainly smaller than the error due to the limited integration time. (Because of the low net fluxes and long radiative time scales in the deep atmosphere, the time it takes for 3D GCM simulations to converge in the deep atmosphere far exceeds simulation runtimes feasible with state-of-the-art computational facilities. This problem is not unique to our model, but universal to 3D GCMs of hot Jupiters when realistic radiative transfer is included.)

Our approximation of a constant CH$_4$/CO ratio also breaks down at pressures lower than $\sim 10^{-4}$ bars, as on the day side CH$_4$ is destroyed photochemically in these regions \citep{MosesEtAl2011,AgundezEtAl2014}. Again, this process is not expected to affect the broadband emission spectra that we are interested in or the temperatures in the regions probed by broadband emission.

The quenched CH$_4$/CO ratio can in general be constrained by comparing the chemical time scale and the mixing time scale. In a 1D picture, the quenched abundances are approximately given by the abundances at the point at which both time scales are equal. In practice, however, the vertical mixing time scale in 1D and 2D models depends on the assumptions about the strength of vertical mixing (parametrized through the eddy diffusion coefficient $K_{zz}$) and the chemical time scale depends on the reaction rates, some of which are not well-known in the pressure and temperature range encountered in hot Jupiter atmospheres. A variety of different 1D and pseudo-2D thermochemical kinetics models of HD 189733b find CH$_4$/CO ratios roughly between 0.001 and 0.5 \citep{MosesEtAl2011,VisscherMoses2011,AgundezEtAl2014,DrummondEtAl2016}.
However, CH$_4$/CO ratios larger than one might also be possible if vertical mixing is very strong \citep{TsaiEtAl2017}.
  GCM simulations with simplified chemical schemes have focused on the hotter HD 209458b and find CH$_4$/CO ratios $\sim$ 0.01 \citep{CooperShowman2006,DrummondEtAl2018}. For HD 189733b, \citet{DrummondEtAl2018HD189733b} find CH$_4$/CO ratios between 0.1 and 0.2.
 We thus choose to treat the CH$_4$/CO ratio as a free parameter and perform 8 simulations varying this ratio from 0.001 to 100 as well as a reference simulation assuming equilibrium chemistry. Although kinetics models favor CH$_4$/CO ratios $<1$, we include the full range for completeness. 
 
\subsubsection{Water and CO$_2$ abundances}
The net reaction that converts CH$_4$ to CO and vice versa is
\begin{equation}
\text{CH}_4+\text{H}_2\text{O} \rightleftharpoons \text{CO} + 3 \text{H}_2.
\label{eq:ch4coreaction}
\end{equation}
Together with the CH$_4$ and CO abundances, the water abundance thus changes as well.  CO and H$_2$O are the only major oxygen-bearing species in the region where quenching dominates the abundances \citep{MosesEtAl2011}, and the total number of oxygen atoms has to be preserved. As a consequence, when reaction (\ref{eq:ch4coreaction}) is quenched, the water abundance is frozen to a constant value as well. This value is directly tied to the CH$_4$ and CO abundances \citep[e.g.,][Fig. 4, 5 and 8]{MosesEtAl2011}. As long as either CO or CH$_4$ dominates over the other, a change in the CH$_4$/CO ratio results in only a marginal change in the water abundance. However, when transitioning from the CO-dominated to the CH$_4$ dominated regime, the water abundance varies by a factor of $\approx2$. Assuming equilibrium chemistry, the water abundance thus varies by this factor of $\approx2$ between the day- and night side of HD 189733b, as can be seen in Figure \ref{fig:h2oabundance}. Including horizontal and vertical transport, however, the water abundance is expected to remain constant between day- and night side due to quenching \citep[Fig. 12]{AgundezEtAl2014}. Since water is such an important infrared absorber, it is necessary to adjust the water abundance in our model to reflect this. We achieve this by adjusting the water abundance such that the total number of oxygen atoms present in water and CO is conserved. The C/H and O/H ratios thus remain at solar values, consistent with our general assumption of solar elemental abundances. 

We leave the abundance of CO$_2$ unchanged from its equilibrium chemistry value, as it is not the dominant carbon species in either equilibrium or disequilibrium chemistry and deviations from equilibrium chemistry are expected to remain moderate compared to CH$_4$ \citep{AgundezEtAl2014}.

\subsubsection{Changes in Opacity}
To illustrate the changes in opacity, the resulting near-infrared opacities for different CH$_4$/CO ratios at a typical photospheric pressure and temperature are shown in Figure \ref{fig:opacities}. The plotted opacities represent the bin-averaged opacity for each of the 196 wavelength bins used for the postprocessing averaged over the inverse of the opacity. 
This kind of average is a good way of representing the effects on the broadband emission. Note that the average is only used for illustration in the figure---in the radiative transfer in the GCM and for post-processing we use the correlated-k method instead (with 11 and 196 bins, respectively). It is obvious from Figure \ref{fig:opacities} that for a higher CH$_4$/CO ratio, the opacity significantly increases for almost all wavelengths in the near and mid-IR. This is mainly because CH$_4$ has many broad absorption bands in the near- and mid-infrared while CO only has  a few narrow absorption bands.  On top of that, with increased CH$_4$, the water abundance also increases by a factor of $\sim 2$. Similar to CH$_4$, water has many broad absorption bands. The only spectral region in which the opacity decreases for high CH$_4$/CO ratios is near the CO absorption band centered around 4.7 $\mu$m. Even in this region, the opacity decreases only by a factor of $\sim$2, much less than it increases at most other wavelengths. This is likely because the water feature centered around 6.3 $\mu$m contributes significantly to the total opacity in this region. Since for every CO molecule that is removed, a water molecule is added, these two changes in opacity partially offset each other in this region.

\section{Results}
\subsection{Reference simulation: equilibrium chemistry case}
First, we give a brief overview of the reference simulation, which assumes equilibrium chemistry. The left column of Figure \ref{fig:temp_composition} shows the temperature and horizontal wind velocities at pressures of 1 mbar, 30 mbars and 1 bar. The pressure levels in the figure are chosen such that the lowest pressure level shown (1 mbar, uppermost panel) is slightly above the photosphere at most infrared wavelengths, while the highest pressure level (1 bar, lowermost panel) is somewhat below the photosphere at most infrared wavelengths. The temperature and wind patterns are typical for a hot Jupiter: The simulation exhibits a large day-night temperature contrast, especially at low pressures, and a superrotating (eastward) equatorial jet. The hottest point is shifted east with respect to the substellar point. The coldest regions in the simulation are east of the antistellar point at roughly 40$^\circ$ to 50$^\circ$ latitude, at the center of two large gyres.

At low pressures, where the radiative time scale is short compared to the dynamical time scale, the temperature difference between day- and night side is largest. In addition to the jet, there is a strong day-to-night flow. As pressure increases and the radiative timescale becomes longer, the eastward equatorial jet dominates more over the day-to-night flow. At a pressure of 1 bar, the equatorial jet efficiently transports heat from the day side to the night side and longitudinal temperature differences along the equator are small. Winds at mid-latitudes are much smaller than at lower latitudes. The cold spots associated with the night side gyres are still several hundred K colder than the equatorial regions.

The ratio of the equilibrium chemistry abundances of CH$_4$ to CO is plotted in the right column of Figure \ref{fig:temp_composition}. In chemical equilibrium, CH$_4$ is the dominant carbon species at high pressures and low temperatures, while carbon preferentially forms CO at low pressures and high temperatures. On isobars, abundances thus directly follow the temperature pattern. On the day side, CO is the prevailing carbon species and is up to 5 orders of magnitude more abundant than CH$_4$.  At most locations of the night side, the abundances of both species become more even, with CH$_4$ prevailing. Methane abundances peak at the night side cold spots at mid-latitudes. As with temperature gradients, the abundance gradients become much less pronounced deeper in the atmosphere: While the abundance ratios span over 8 orders of magnitude at the 1 mbar level, they differ by only 3 orders of magnitude at the 1 bar level.

\subsection{Simulations with quenched abundances: Thermal structure}
\label{sec:thermalstructure}
\begin{figure*}
\plotone{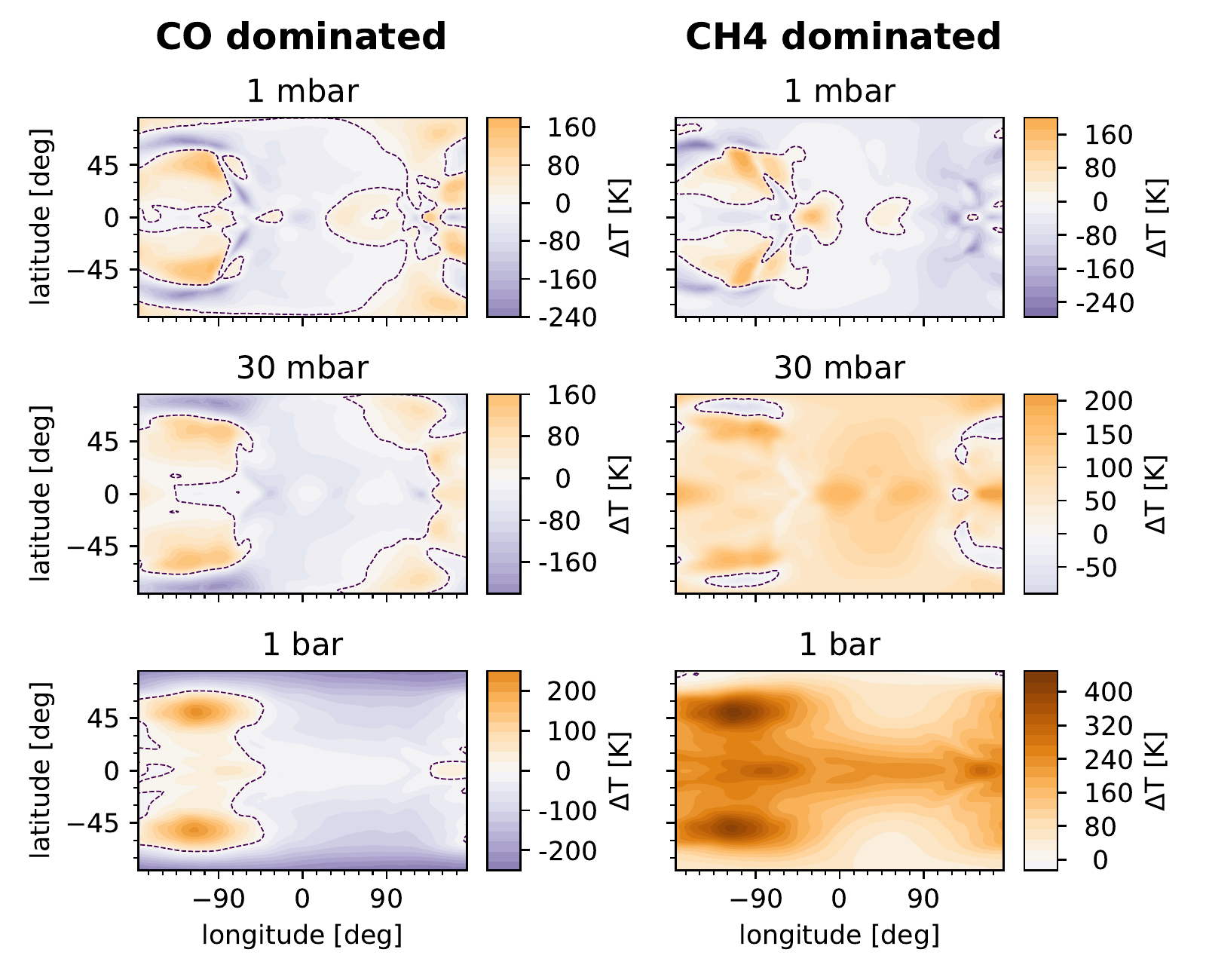}
\caption{Temperature difference of the time-averaged simulation output with respect to the equilibrium chemistry simulation for two simulations with a quenched CH$_4$/CO ratio (left: CH$_4$/CO=0.01, right: CH$_4$/CO=100) at pressures of 1 mbar, 30 mbars and 1 bar. Positive temperature differences (i.e. quenched simulation is hotter) are shown in orange, negative temperature differences are purple. Dashed lines indicate the zero contour.  \label{fig:differences_ch4co}}

\end{figure*}

\begin{figure*}
\plotone{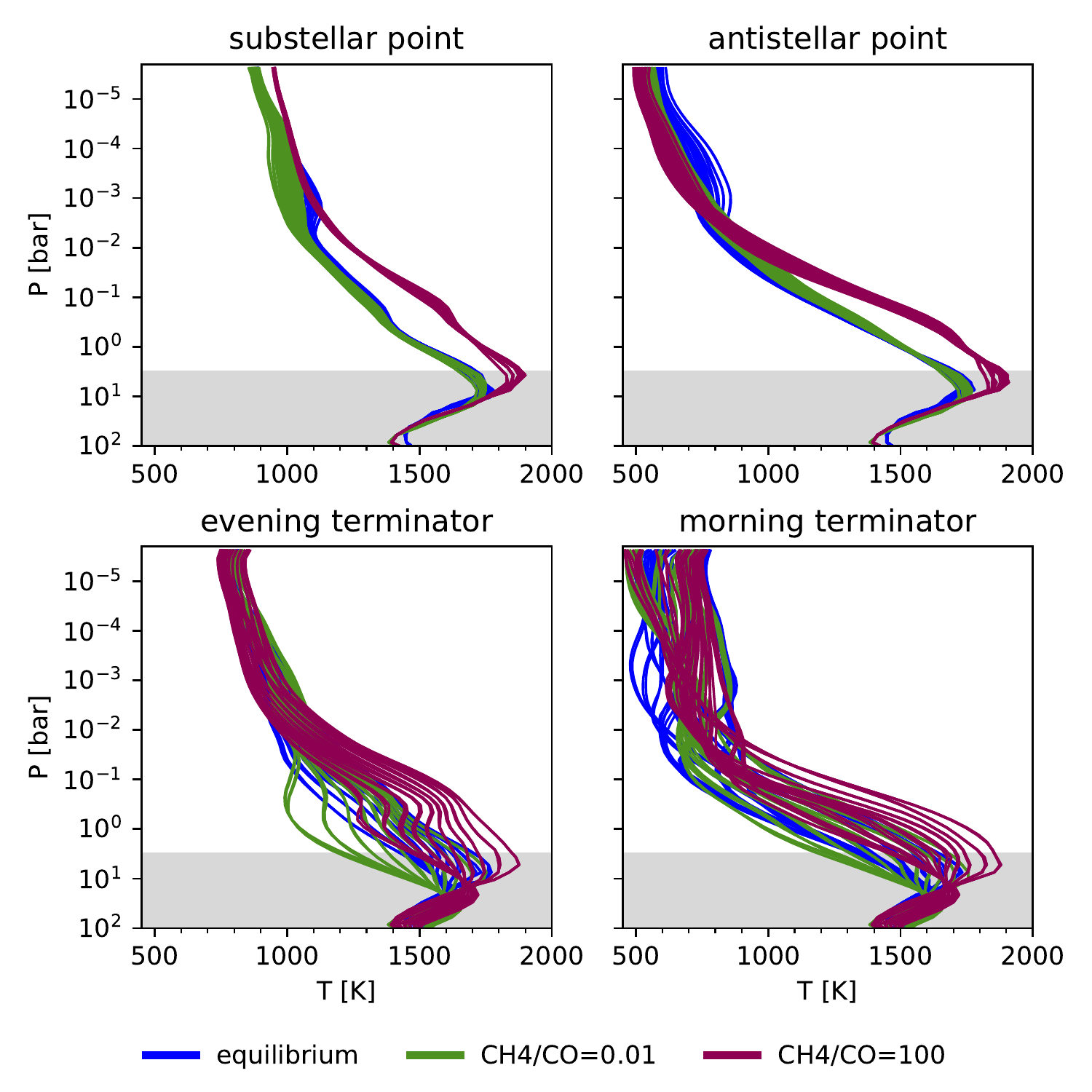}
\caption{Pressure-temperature profiles. 
The equilibrium chemistry simulation is shown in blue, the CH$_4$/CO=0.01 simulation in green and the CH$_4$/CO=100 simulation in magenta. The shaded grey region indicates pressure ranges for which the simulation has not converged due to the long radiative timescale in deep layers of the atmosphere. Coincidentally, our assumption of a constant quenched CH$_4$/CO ratio breaks down at roughly a similar pressure. Upper left: Profiles within 10$^\circ$ of the substellar point. Upper right: Profiles within 10$^\circ$ of the antistellar point. Lower left: Profiles within $\pm 2 ^\circ$ of the evening terminator. Lower right: Profiles within $\pm 2 ^\circ$ of the morning terminator. The temperature profiles in this figure are available as Data behind the Figure. \label{fig:ptprofiles}
} 
\end{figure*}

In response to the different opacities, the thermal structure changes in the simulations with disequilibrium chemistry compared to the reference case. We consider two of the quenched simulations, CH$_4$/CO=0.01 and CH$_4$/CO=100, in detail to illustrate these changes for the CO dominated and the CH$_4$ dominated regime. Figure \ref{fig:differences_ch4co} shows the temperature difference with respect to the reference case on isobars. Vertical pressure-temperature profiles from these two simulations along with the reference simulation are plotted in Figure \ref{fig:ptprofiles}.

In the CO dominated regime (the regime favored by kinetics models), the day side is cooler compared to the equilibrium case, while the temperature of large parts of the night side, including the midlatitude cold spots, increases. Temperatures also drop at high latitudes, especially on the night side. All of these effects can be seen in the left column of Figure \ref{fig:differences_ch4co}. In the particular case shown (CH$_4$/CO=0.01), day side temperatures drop by about 50 K throughout most of the photosphere. This is also evident in the pressure-temperature profiles near the substellar point (upper left panel of Figure 4). A closer look at the pressure-temperature profiles reveals that the vertical temperature gradient is slightly larger compared to the equilibrium chemistry case near the substellar point, but slightly smaller at the antistellar point. This is likely because on the day side, greenhouse gases (CH$_4$, H$_2$O) are added compared to the equilibrium composition while on the night side greenhouse gases are taken away. The  CH$_4$/CO=0.001 simulation (not shown) displays very similar changes in thermal structure, but with a somewhat larger amplitude.

Our interpretation for why the day side is cooler is related to energy-balance: 
The changes in opacity lead to only slight changes in albedo, so for all the CH4/CO ratios considered, the total amount of absorbed starlight is nearly the same.  Since the simulations achieve an approximate energy balance (energy gained equals energy lost), this means that the total IR energy radiated to space (from the entire planet) is also nearly the same regardless of the CH$_4$/CO ratio.
Compared to equilibrium chemistry, the overall infrared opacity decreases on the night side. Therefore, the radiation on the night side escapes to space from deeper levels, where the atmosphere tends to be hotter, leading to a larger flux being radiated away. As a consequence, less energy is returned to the day side and the day side cools, which in turn leads to a dayside radiating less IR flux.

In the (less likely) CH$_4$ dominated case, the temperature changes are more striking:  Below roughly the 10 mbar level, temperatures are significantly hotter at almost all longitudes and latitudes.  In the CH$_4$/CO=100 simulation, the temperature difference with respect to the reference simulation reaches 200 to 400 K at the 1 bar level in the equatorial region (see right column of Figure \ref{fig:differences_ch4co}). In this simulation, the location of the night side cold spots shifts to slightly higher latitudes. In Figure \ref{fig:differences_ch4co}, this is visible in the center right panel as the light blue regions  and  in the upper right panel as the two small deep purple regions at $\approx \pm 60^\circ$ latitude in conjunction with two orange regions at lower latitudes.

At the 1 mbar level, temperatures are similar or slightly colder compared to the reference simulation. Looking at the temperature profiles in Figure \ref{fig:ptprofiles}, it is obvious that the temperature gradient between 1 mbar and 1 bar is steeper in the CH$_4$ dominated case than for the other simulations at all four locations shown, signifying a stronger greenhouse effect. Given that in the CH$_4$ dominated case the abundances of the greenhouse gases methane and water are enhanced compared to equilibrium chemistry by several orders of magnitude and a factor of $\sim2$, respectively, at any location other than the night side mid-latitude cold spots, this behavior is expected.

\subsection{Phase Curve Predictions and Comparison with Observations}
\label{sec:phasecurvepredictions}
\begin{figure*}
\plotone{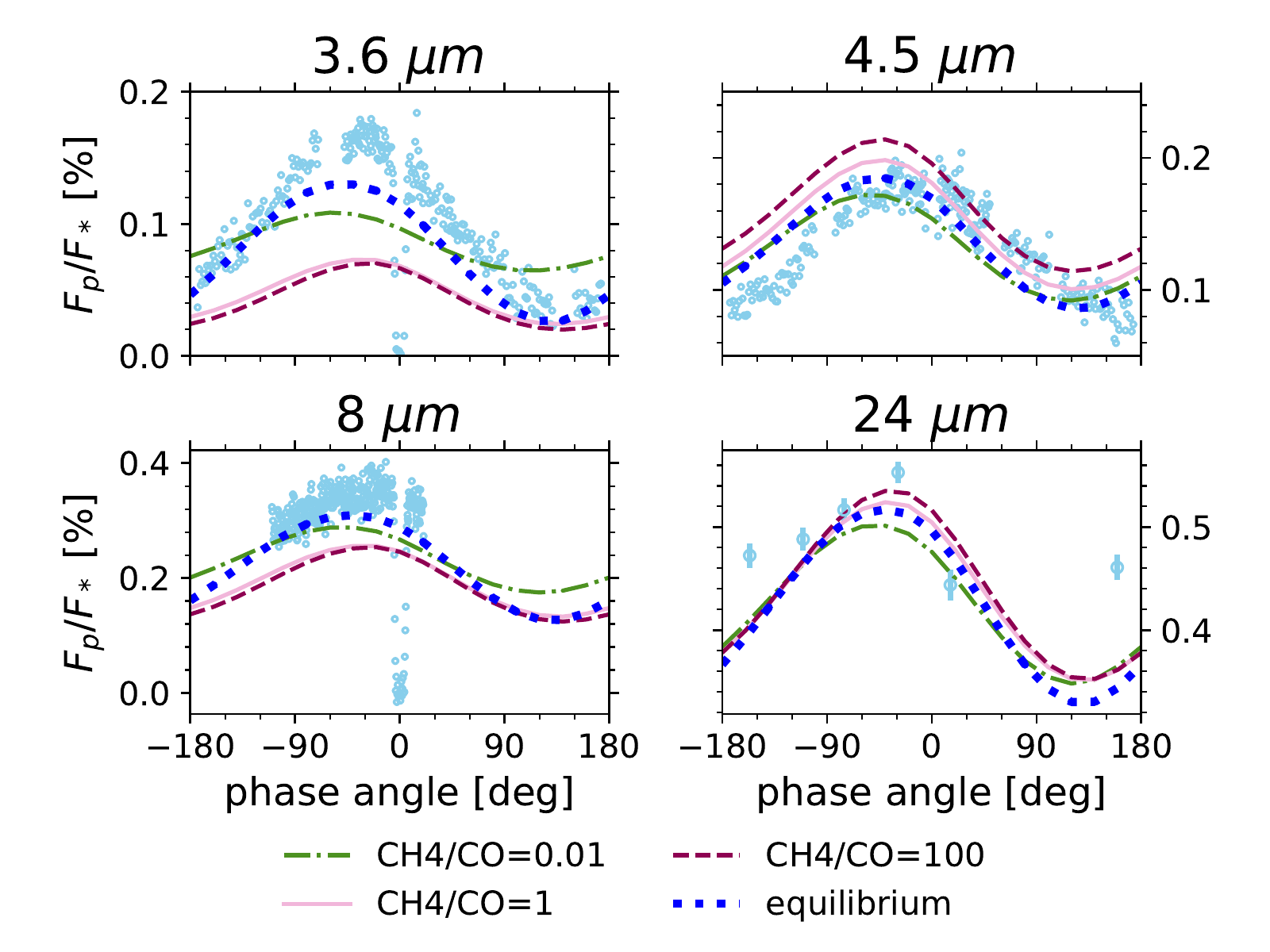}
\caption{Phase curves predicted from our models and available observational data in the Spitzer 3.6, 4.5, 8 and 24 micron bands. The green dash-dotted line represents the CH$_4$/CO=0.01 simulation, the light pink solid line CH$_4$/CO=1, the magenta dashed line CH$_4$/CO=100 and the thick dotted blue  line the reference simulation. The observational data (light blue circles) are the same as in Fig. 12 in \cite{KnutsonEtAl2012} and are taken from \cite{KnutsonEtAl2012} (3.6 and 4.5 $\mu$m), \cite{KnutsonEtAl2007,KnutsonEtAl2009_HD189,AgolEtAl2010} (8 $\mu$m) and \cite{KnutsonEtAl2009_HD189} (24 $\mu$m). All model phase curves in this figure are available as Data behind the Figure. \label{fig:phasecurves_with_data}
} 
\end{figure*}

We post-process the GCM results as described in Section \ref{subsec:radtran} to obtain phase curves and emission spectra and compare them to the available observational data (Figure \ref{fig:phasecurves_with_data}). In short, the fit with respect to the observational phase curve worsens substantially in the 3.6 $\mu$m band for all quenched ratios while the changes in the 4.5 $\mu$m band are negligible in the expected CO dominated case and lead to a worse fit in the CH$_4$ dominated regime.

In the 3.6 $\mu$m band, the phase curve amplitude decreases for all quenched simulations compared to equilibrium chemistry. A similar, but less drastic behavior can be seen in the 8 $\mu$m band. This behavior is expected for wavelength regions with strong methane absorption bands, including these two band passes: In equilibrium chemistry, the outgoing radiation on the day side, where there is little methane, probes deeper, hotter layers while on the night side, where more methane is present, the radiation is emitted from higher, cooler layers. Thus the temperature varies more strongly from day to night on the photosphere in these bands (which differs in pressure from day to night) than it does on isobars. This enhances the day-night contrast of the phase curve relative to what would occur if the photosphere were at constant pressure \citep[see also][]{Dobbs-DixonCowan2017}. When assuming a constant methane abundance throughout the atmosphere, however, regardless of the exact value, the outgoing radiation emerges from similar pressure regions everywhere on the planet and the phase curve amplitude decreases. Consistent with this picture, the day side fluxes from the CH$_4$/CO=0.01 simulation are relatively close to those from the equilibrium chemistry simulation, while the night side fluxes from the CH$_4$ dominated simulations match those from equilibrium chemistry at the flux minimum.

In the 4.5 $\mu$m band, the night side fluxes are similar to or higher than the value from the equilibrium chemistry simulation for all quenched values. At first, this seems surprising, given that \cite{KnutsonEtAl2012} base their argument for disequilibrium chemistry almost entirely on the fact that equilibrium chemistry models over-predict the 4.5 $\mu$m night side fluxes, arguing that adding CO on the night side will shift the photosphere in this band to a higher, colder atmospheric layer and thus decrease the flux. However, they ignore that when changing the CO abundance through quenching, the water abundance should also change. Water has significant opacity in the 4.5 micron band as well. Averaged across the band, its molecular absorption cross section is comparable to that of CO (see e.g. Figure 5  in \citet{FortneyEtAl2006} or Figures 2 and 3 in \citet{SharpBurrows2007}). For each CO molecule added, an H$_2$O molecule is removed (assuming metallicity is held constant), and the resulting changes in opacity in the 4.5 micron band due to CO and H$_2$O largely offset each other (see also Section \ref{subsec:methods_disequilibriumchemistry} and Figure \ref{fig:opacities}). The changes in the 4.5 micron phase curve are thus mainly due to the change in thermal structure and not due to a change of the photospheric level. In the hotter CH$_4$ dominated case, the flux increases at all phase angles in this band. In the CO dominated case, the flux near the phase curve maximum decreases slightly compared to the equilibrium chemistry case, reflecting the cooler day side, while the phase curve follows closely the equilibrium chemistry phase curve at other phase angles.

In the 24 $\mu$m band, the difference between the disequilibrium and equilibrium chemistry simulations is relatively small. None of the predicted phase curves match the available data on the night side well. However, as observations in this band do not cover all phase angles, the shape of the phase curve in this band remains uncertain.

\subsection{Phase Curve Trends}
We now examine trends in the phase curve properties with the quenched ratio. Two factors control how the phase curve in the quenched case differs from the the equilibrium chemistry case. First, the change in opacities directly impacts the thermal structure of the planet, as discussed in Section \ref{sec:thermalstructure}. Second, the change in opacities causes the photospheric level to shift upwards or downwards. As a result, the outgoing radiation probes cooler or hotter layers. This change in the location of the photosphere can also be studied by postprocessing the output of GCM simulations assuming equilibrium chemistry with quenched opacities instead of self-consistently including quenched opacities in the GCM. 
It is thus of significant interest to examine the relative importance of both of these factors in shaping the predicted phase curve. If it turned out that the change in the thermal structure had only a small effect on the phase curve, modelers could restrict themselves to including disequilibrium chemistry in the postprocessing of the GCM output, which is simpler and computationally less expensive. In addition to computing the phase curves from the simulations with disequilibrium chemistry (self-consistent opacities), we thus also computed phase curves by postprocessing the output from the equilibrium chemistry simulation with quenched opacities.

\begin{figure*}
\plotone{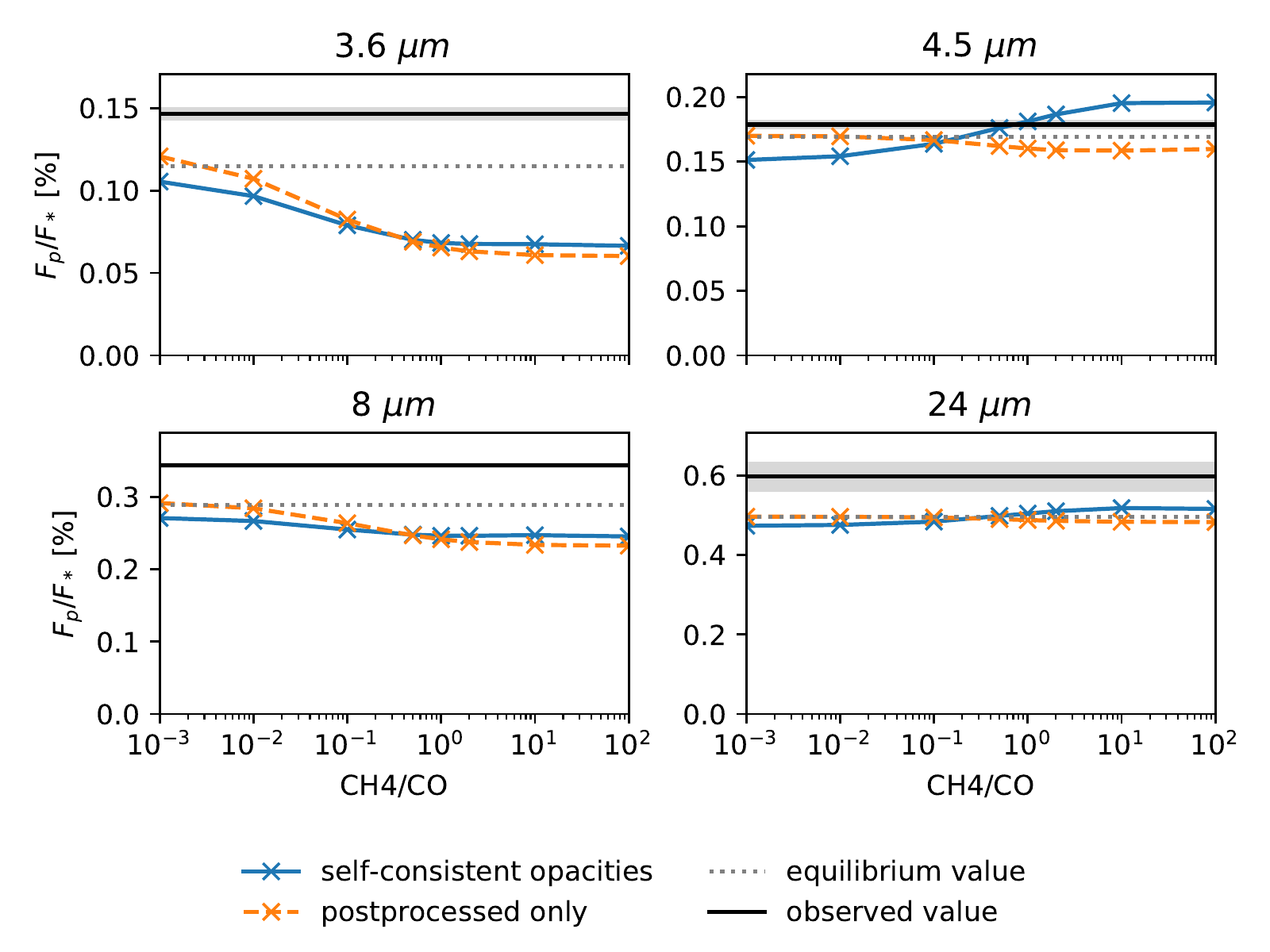}
\caption{Secondary eclipse depths for different quenched CH$_4$/CO ratios. Values from the GCM simulations including the quenched opacities are shown as blue crosses connected by a solid line, while values obtained from postprocessing the equilibrium chemistry simulation are shown as orange crosses connected by a dashed line. The equilibrium value is indicated by the dashed grey line and the observed values from \cite{KnutsonEtAl2012} are shown as a solid black line with the 1$\sigma$ errors shaded in grey.  \label{fig:secondary_eclipse_bands}
} 
\end{figure*}

\begin{figure*}
\plotone{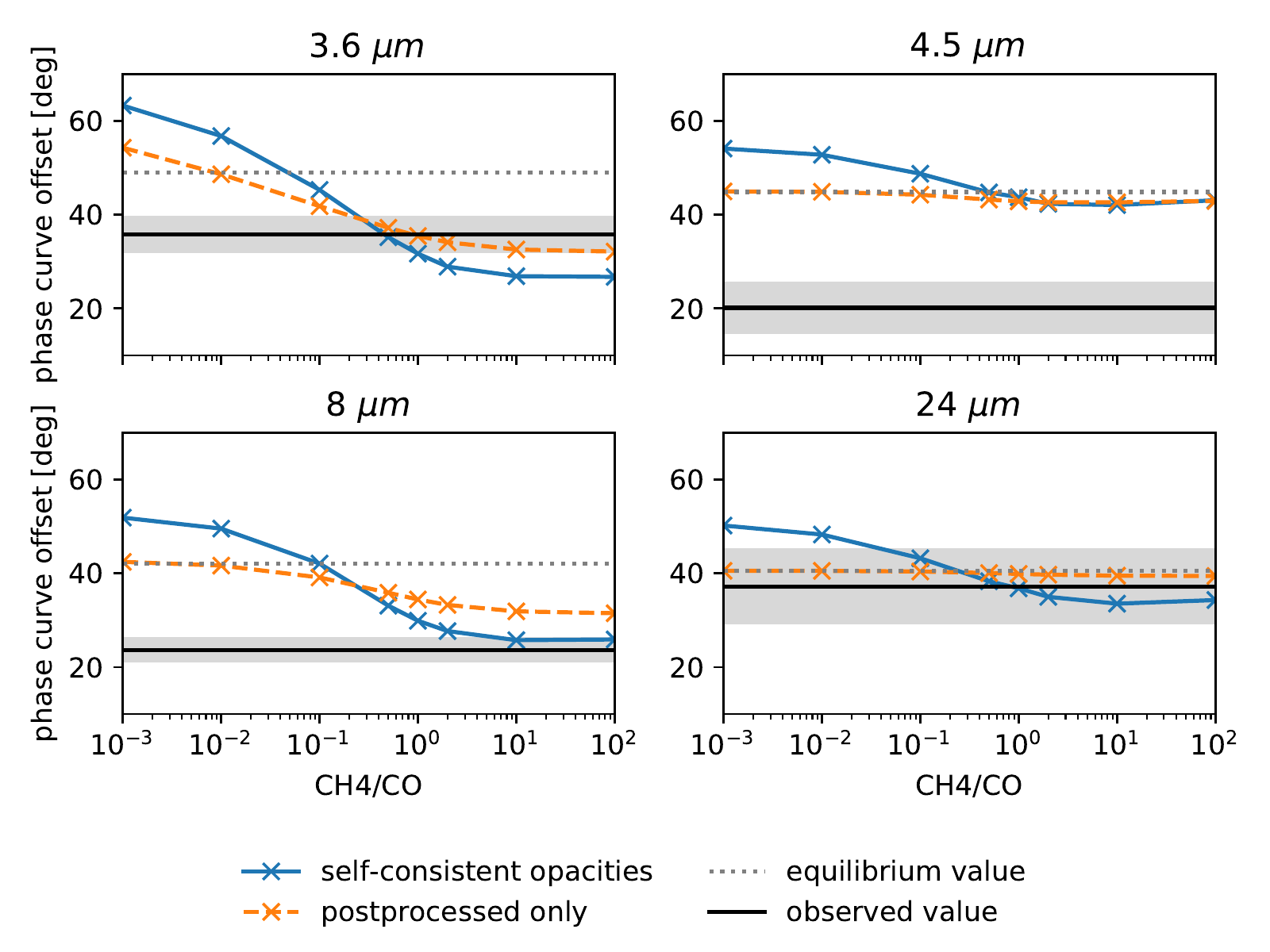}
\caption{Eastward offsets of the maximum flux with respect to secondary eclipse for different quenched CH$_4$/CO ratios. Values from the GCM simulations including the quenched opacities are shown as blue crosses connected by a solid line, while values obtained from postprocessing the equilibrium chemistry simulation are shown as orange crosses connected by a dashed line. The equilibrium value is indicated by the dashed grey line and the observed values from \cite{KnutsonEtAl2012} are shown as a solid black line with the 1$\sigma$ errors shaded in grey.  \label{fig:phasecurve_offset}} 
\end{figure*}

\begin{figure*}
\plotone{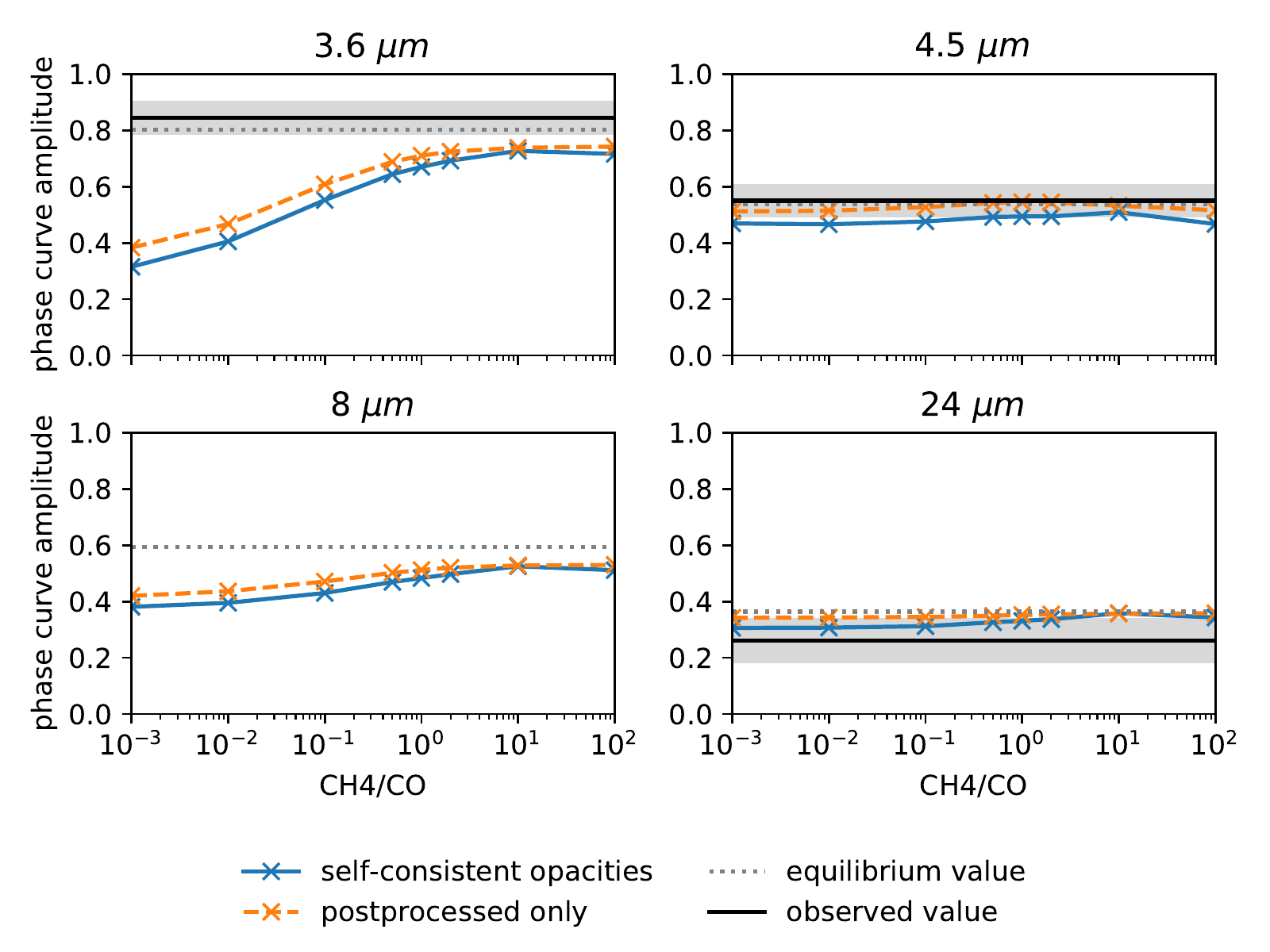}
\caption{Normalized phase curve amplitudes, defined as $A=1-F_{min}/F_{max}$,  for different quenched CH$_4$/CO ratios. Values from the GCM simulations including the quenched opactities are shown as blue crosses connected by a solid line, while values obtained from postprocessing the equilibrium chemistry simulation are shown as orange crosses connected by a dashed line. The equilibrium value is indicated by the dashed grey line and the observed values from \cite{KnutsonEtAl2012} are shown as a solid black line with the 1$\sigma$ errors shaded in grey. Note that for the 8 $\mu m$ band there is no observed value, as observations in this band did not cover the night side. \label{fig:phasecurve_amplitude_bands}} 
\end{figure*}

Figures \ref{fig:secondary_eclipse_bands} to \ref{fig:phasecurve_amplitude_bands} show trends in secondary eclipse depth, phase curve offset and phase curve amplitude, respectively, with the quenched ratio for phase curves obtained from both simulations with self-consistent opacities and from the equilibrium chemistry simulation postprocessed with quenched opacities. If both curves align, the change is dominated by the shift in the photosphere pressure. If there are large discrepancies between the postprocessed and the self-consistent GCM phase curves, the change in thermal structure plays an important role.

In the 3.6 and 8 $\mu$m bands, which include strong methane absorption features, the phase curve amplitude increases with increasing methane abundance (Figure \ref{fig:phasecurve_amplitude_bands}), while the phase curve offset and the secondary eclipse flux decrease (Figures \ref{fig:phasecurve_offset} and \ref{fig:secondary_eclipse_bands}, respectively). This is expected as the photosphere in these bands shifts to higher levels of the atmosphere with stronger day-night contrasts. All of these trends flatten once CH$_4$ starts to dominate over CO. The phase curve amplitudes from the post-processed simulations closely match the ones with self-consistent opacities, indicating that the change of the photospheric level is the dominant cause for this trend. For the phase curve offset, the direction of the trend is similar; however, the trend is weaker in the post-processed only points, indicating that the change in thermal structure significantly contributes to the trend. 

In the 4.5 $\mu$m band, secondary eclipse flux, phase curve offset and  phase curve amplitude from the postprocessed-only simulations are close to the equilibrium value, indicating that the location of the photosphere changes only marginally. In the simulations with self-consistent opacities, however, the secondary eclipse flux is increasing with increased methane abundance, in line with the hotter temperatures in the simulation described in Section \ref{sec:thermalstructure}. The phase curve offset also decreases with increasing CH$_4$ in the CO dominated regime. The phase curve amplitudes stay similar.

In the 24 $\mu$m band, there are no trends in secondary eclipse and phase curve amplitude. For the phase curve offset,  the simulations with self-consistent opacities replicate the same decreasing trend as in other wavelength bands, as would be expected for a trend due to a change in thermal structure.

It is also of interest to examine how the described trends compare to the properties of the observed phase curve (indicated in black with grey error bars in Figures \ref{fig:secondary_eclipse_bands} to \ref{fig:phasecurve_amplitude_bands}). In the 3.6 $\mu$m and 8 $\mu$m bands, the secondary eclipse depth is closer to the observed value for low CH$_4$/CO values than for high CH$_4$/CO values. The phase curve offset, in contrast, matches the observed value for CH$_4$/CO values near 0.5 in the 3.6 $\mu$m band and is closest to the observed value for CH$_4$/CO values greater than one for the 4.5 $\mu$m and 8 $\mu$m bands. The phase curve amplitude in the 3.6 $\mu$m band is closer to the observed value for high CH$_4$/CO ratios. In short, among the quenched simulations there is no one CH$_4$/CO ratio that matches the phase curve properties best, as low CH$_4$/CO ratios match the secondary eclipse better but high CH$_4$/CO ratios match the phase curve offset better. This further underlines our finding from Section \ref{sec:phasecurvepredictions} that quenching of CH$_4$ and CO does not provide a good explanation for the shape of the observed phase curves.

\subsection{Emission spectra}
\begin{figure*}
\plotone{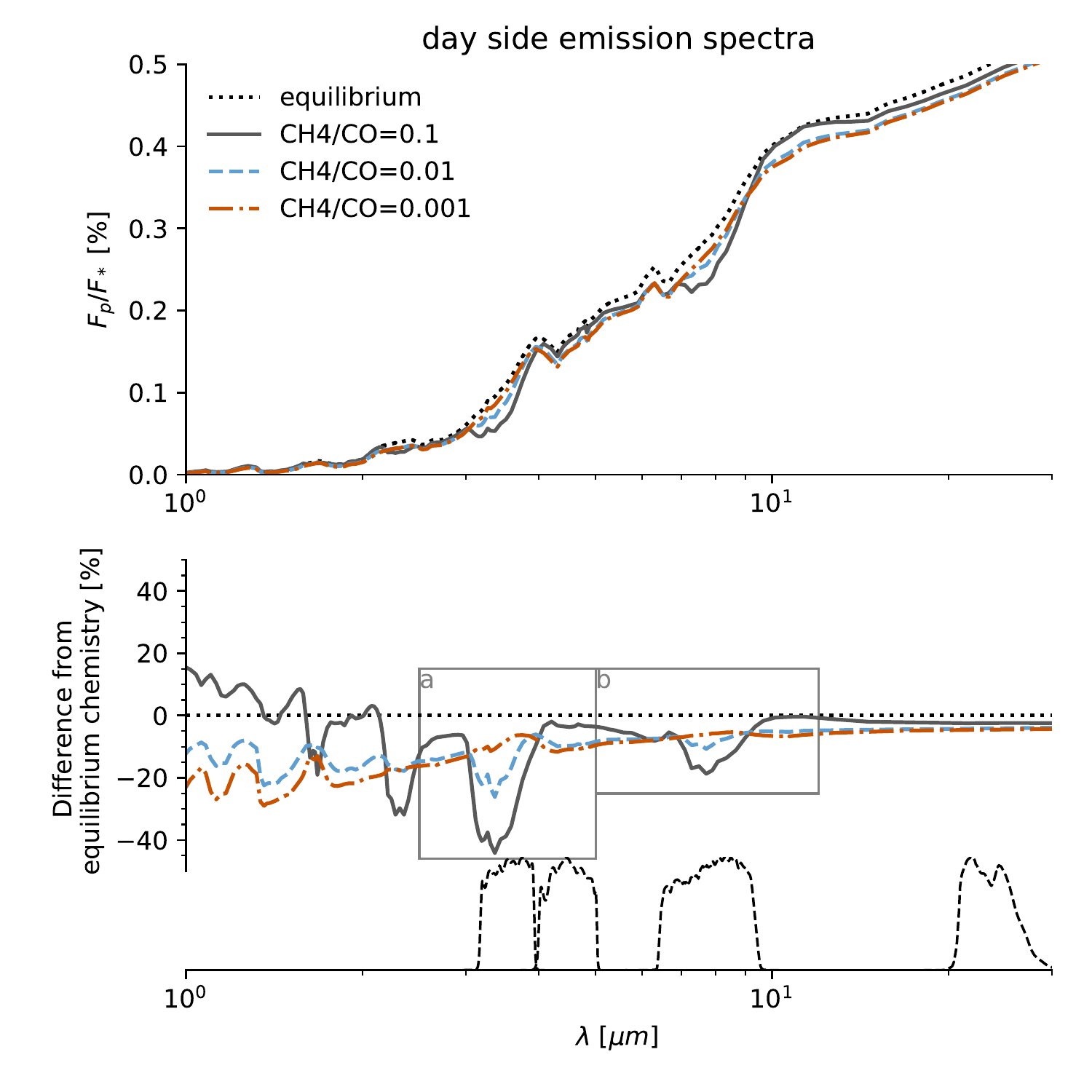}
\caption{Predicted day side emission spectra, as would be observed at secondary eclipse. The ratio of the planet flux to the stellar flux is plotted in the top panel. The bottom panel shows the relative difference of the flux in each of the quenched simulations to the flux in the equilibrium chemistry simulation. The regions plotted in panels a and b of Fig. \ref{fig:spectra_jwst} are indicated with gray boxes. The dashed black lines at the bottom indicate the filter sensitivity profiles 
of the Spitzer 3.6 $\mu$m, 4.5 $\mu$m, 8 $\mu$m and 24 $\mu$m bands. The emission spectra in this Figure are available as Data behind the Figure. \label{fig:spectra_day}}
\end{figure*}

\begin{figure*}
\plotone{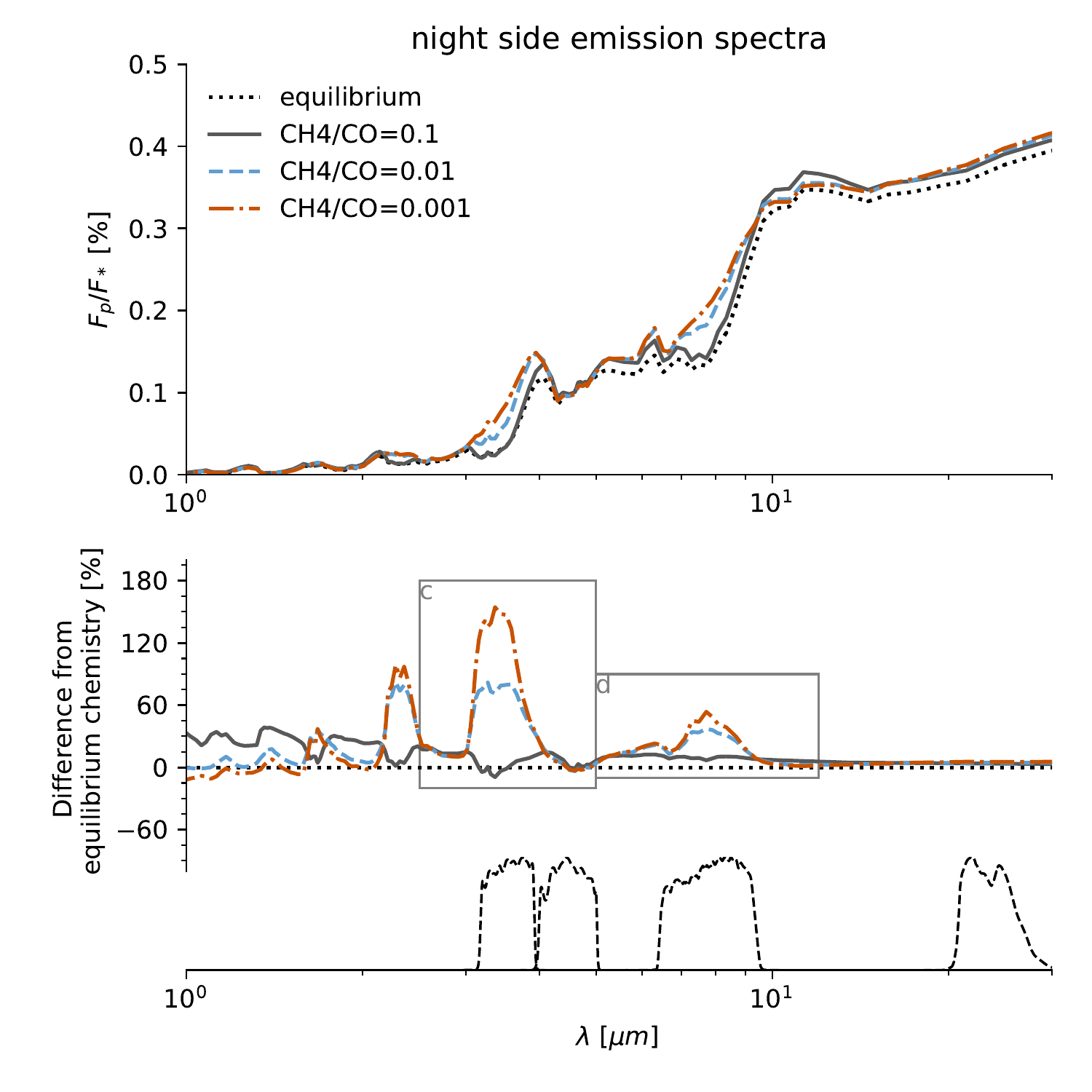}
\caption{Same as Fig. \ref{fig:spectra_day} but for night side emission spectra, as would be observed right before or after transit. The emission spectra in this figure are available as Data behind the Figure.\label{fig:spectra_night}}
\end{figure*}

To move beyond a discussion restricted to the Spitzer bands, we plot the predicted emission spectra of the day side (Figure \ref{fig:spectra_day}) and night side (Figure \ref{fig:spectra_night}) for the CO-dominated simulations. We focus on our CH$_4$/CO $<$ 1 cases as these are expected to be more likely for the temperature range of HD 189733b. Even though we found in the previous subsections that the disequilibrium chemistry cases from our model do not match existing observations, it is instructive to look at the effect of disequilibrium chemistry on the emission spectra. The findings could also be applied to other planets in a similar temperature regime.  

In general, the day side fluxes from the quenched simulations are lower than the fluxes from the reference case, while the night side fluxes from the quenched simulations are significantly higher than for the reference case. This is consistent with the change in energy balance mentioned in Section \ref{sec:thermalstructure}.

The quenched day side emission spectra deviate moderately from the equilibrium chemistry spectrum. The difference is largest in the water absorption bands between 1 and 2 $\mu$m for the CH$_4$/CO=0.001 spectrum (up to -30\%) and largest near the 3.3 $\mu$m methane absorption band for the CH$_4$/CO=0.01 and 0.1 spectra (up to -25\% and -40\%, respectively). In the CH$_4$/CO=0.1 spectrum, methane absorption bands clearly show up.

In the night side spectra, for CH$_4$/CO=0.001 and 0.01 there are large differences compared to the equilibrium chemistry case, especially in the methane absorption bands. The largest flux difference can be found near the 3.3 $\mu$m methane band (up to 160\% and 80\%, respectively), but the 2.3 $\mu$m (up to 100\% and 80\%) and 7.7 $\mu$m (up to 55\% and 40\%) methane bands also show substantial flux differences. For the CH$_4$/CO=0.1 case, the difference to equilibrium chemistry is in general much smaller and the largest difference can be seen in the water absorption bands between 1 and 2 $\mu$m. In the region around 4.5 $\mu$m, the differences from the equilibrium chemistry case are small for all three quenched ratios, reinforcing our previous conclusion that this wavelength region is not suitable for detecting disequilibrium chemistry on the night sides of hot Jupiters.

The fact that disequilibrium chemistry mainly affects spectral regions with CH$_4$ bands, as well as H$_2$O bands, raises the question how the spectral signatures of disequilibrium chemistry can be distinguished from the spectral signatures of increased or decreased CH$_4$ and H$_2$O abundances in equilibrium chemistry due to a non-solar C/O or C/H ratio. While for a single secondary eclipse or night side emission spectrum, these scenarios may look similar, we expect that the combination of a secondary eclipse and night side emission spectrum would be able to discriminate between these scenarios.

\subsection{Simulated JWST spectra}
\begin{figure*}
\plotone{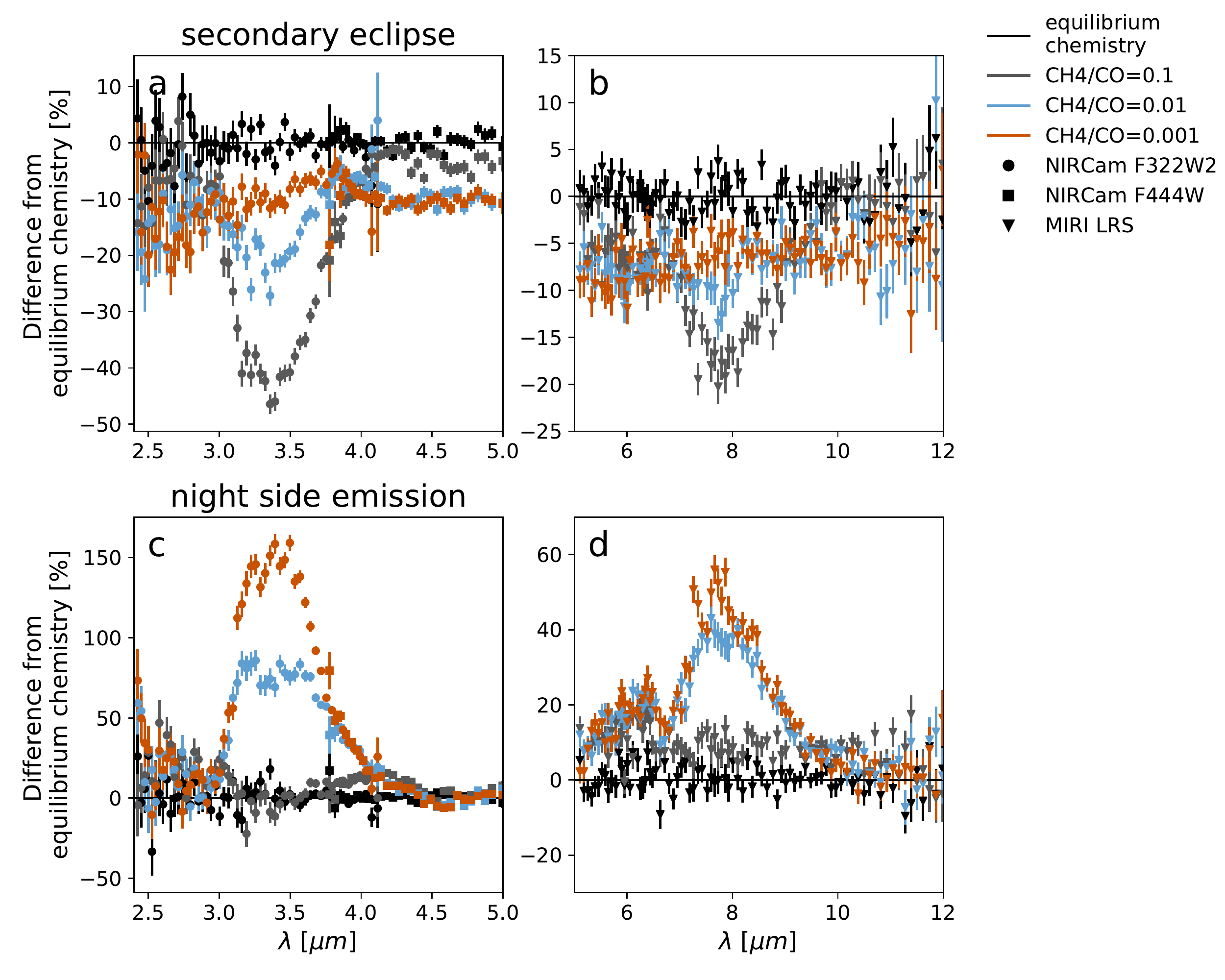}
\caption{Simulated JWST observations for NIRCam (left panels) using the F322W2 (circles) and F444W (squares) grisms and MIRI LRS (right panels, triangles). Plotted is the difference to the flux predicted by the equilibrium chemistry model, binned to a spectral resolution of $R\approx50$, with 1$\sigma$ error bars. The simulated observations in this figure are available as Data behind the Figure. \label{fig:spectra_jwst}}
\end{figure*}
In order to determine whether the CH$_4$/CO ratio can be distinguished with future observations, we simulated JWST observations of HD 189733b using PandExo \citep{BatalhaEtAl2017}. PandExo is a noise simulator that uses the Space Telescope Science Institute's exposure time calculator, Pandeia \citep{PontoppidanEtAl2016}. Using PandExo, we find that in many NIRSpec and NIRISS observing modes the detector becomes saturated due to HD 189733b's brightness (J=6.07). We therefore simulate data for two of NIRCam's grisms, F322W2 (2.413-4.083 $\mu$m) and F444W (3.835-5.084 $\mu$m), and MIRI Low-Resolution Spectroscopy (LRS) (5-12 $\mu$m). These three modes all remain below 80\% full well saturation across the detector. The NIRCam grisms ultilize the 64 x 2048 pixel subarray to reduce readout times. MIRI LRS uses a dedicated slitless prism subarray region.

Each instrument mode was simulated assuming only a single eclipse with equal time in eclipse as out of eclipse. We also assume a 40 PPM noise floor, which is similar to the noise floors found in WFC3 on HST \citep{KreidbergEtAl2014}. We simulated data for different CH$_4$/CO ratios and for equilibrium chemistry. Observations were simulated for both the day side and the night side. Since PandExo only models eclipses and transits and not phase curves, the planet's night side spectrum was given to PandExo as if it were the planet's day side since the expected noise should be the same, except for the reduced brightness of the planet's night side. We plot the results for the CO dominated simulations  in Figure \ref{fig:spectra_jwst} in terms of the difference to the equilibrium chemistry model. We also include the simulated observations for the equilibrium chemistry case to give a sense of the expected residuals.
Note that the scale on the y-axis is different in each panel. For better orientation, the regions shown in each panel are indicated with grey boxes in Figures \ref{fig:spectra_day} and \ref{fig:spectra_night}.

In secondary eclipse, all quenched cases can be distinguished from the disequilibrium chemistry case in all three observing modes. The different CH$_4$/CO ratios 0.1, 0.01 and 0.001 can also clearly be distinguished from each other. With MIRI LRS, the fluxes in the CH$_4$/CO=0.01 and 0.001 cases are relatively close to each other, but can still be distinguished from each other at  greater than $3\sigma$ confidence.

Looking at night side emission spectra, the relative deviations from the equilibrium chemistry spectra are much larger. The CH$_4$/CO=0.01 and 0.001 can easily be distinguished from equilibrium chemistry in all three observing modes. The night side fluxes of the CH$_4$/CO=0.1 case are in general much closer to the equilibrium chemistry case in the wavelength ranges covered by these observing modes, but still can be distinguished from disequilibrium chemistry. With the NIRCam F322W2 grism and MIRI LRS, it is possible to clearly distinguish the three shown CH$_4$/CO ratios from each other as well. With simulated data from the NIRCam F444W grism, it is still possible to tell the different CH$_4$/CO ratios apart from each other at $3\sigma$ confidence, but differences between the quenched ratios are only evident at wavelengths shorter than 4 $\mu$m, at the very edge of the wavelength range of this grism. Thus, it may not be the first choice for observing disequilibrium chemistry.

Our intent in presenting these simulated observations is to demonstrate the need of including disequilibrium chemistry when interpreting future observations with JWST and to guide observers in which regions to look for signatures of disequilibrium carbon chemistry. However, it would be premature to conclude that one would be able to determine the CH$_4$/CO ratio, as other factors not considered in this work can influence the results of GCM simulations and the emission spectra, including clouds, metallicity, C/O ratio and atmospheric drag (see Section \ref{subsec:discussion_otherfactors}). Especially an optically thick cloud deck on the night side could limit our ability to observe signatures of disequilibrium chemistry, as it would block emission from the layers in the atmosphere below the cloud deck. This could mute the strongly visible CH$_4$ features in Figures \ref{fig:spectra_day} to \ref{fig:spectra_jwst} in which the difference to equilibrium chemistry is most obvious. Future work in this direction is necessary.

\section{Discussion}
\subsection{Importance of including disequilibrium chemistry in GCM}
In their study of HD 209458b, \citet{DrummondEtAl2018} find that the effect of transport-induced disequilibrium chemistry on the temperature structure and winds is $\sim 1\%$ relative to otherwise similar models with chemical equilibrium. In contrast, in our simulations of HD 189733b we find an effect on the temperature of up to $\sim 5-10 \%$ across large regions (larger changes in very localized regions) in the more likely CO-dominated regime and even larger temperature changes in the CH$_4$-dominated regime. While this paper was in review, \citet{DrummondEtAl2018HD189733b} found a similar result in their simulations of HD 189733b.
 This raises the question of why this effect is so much larger on HD 189733b. The answer lies in the cooler temperature of HD 189733b. On the hotter HD 209458b (equilibrium temperature $T_{eq}\approx1450$ K), CO remains the dominant species even on the night side in both equilibrium and disequilibrium chemistry. The change in opacity due to disequilibrium chemistry is thus less drastic. In contrast, on the cooler HD 189733b ($T_{eq}\approx1200$ K) the night side is cold enough that with equilibrium chemistry, CH$_4$ becomes the dominant species on a significant fraction of the night side. In disequilibrium chemistry, however, CO is expected to be the dominant species everywhere at pressures below 1 bar. Thus, when including disequilibrium chemistry in the radiative transfer, for most of the night side the dominant carbon species switches from CH$_4$ to CO, significantly changing the opacities. With this switch between CH$_4$ and CO as dominant species, the H$_2$O abundance changes by a factor of about two as well, further affecting the opacities. For planets much cooler than HD 189733b, even the day side becomes CH$_4$ dominated in both equilibrium and disequilibrium chemistry. Regardless of exact disequilibrium abundances, the dominant species remains the same as in equilibrium chemistry on these planets. For this situation, we would thus again expect a smaller effect on the opacities and thus on the resulting temperatures.

 Therefore, we expect that including the effect of disequilibrium carbon chemistry on the opacities in the GCM is most important for planets for which equilibrium chemistry predicts a CO dominated day side but a CH$_4$ dominated night side. 
Assuming solar metallicity, the equilibrium abundance of CH$_4$ becomes comparable to the CO abundance on the coldest regions of the night side for planets with equilibrium temperatures below $\sim 1300$ K \citep[Fig. 7, taking the morning terminator profiles to be representative of the coldest regions of the planet]{KatariaEtAl2016}. This demarcates the upper boundary of the regime in which we expect disequilibrium carbon abundances to be important in GCMs. The colder end of this regime has been less explored by GCMs. Based on 1D models,  for solar composition the day side is expected to transition from CO-dominated to CH$_4$-dominated roughly at equilibrium temperatures around 700 K, with the transition occurring at slightly higher temperatures in equilibrium chemistry and at slightly lower temperatures in disequilibrium chemistry \citep{MosesEtAl2013GJ436b,VenotEtAl2014,MiguelKaltenegger2014}. We thus expect that it is important to include disequilibrium CH$_4$ and CO abundances in GCMs of hot Jupiters for planets with equilibrium temperatures between $\sim$ 600 K and 1300 K, assuming solar abundances. As the temperature of the transition between CH$_4$ and CO strongly depends on the metallicity and C/O ratio, this temperature range may change depending on atmospheric composition. Future work spanning a range of planets is necessary to test this prediction. This could include both simulations coupling a chemical scheme to a GCM similar to \citet{DrummondEtAl2018} and \citet{CooperShowman2006} and simulations with a simpler approach comparable to ours.

While we expect these findings to qualitatively also apply to atmospheres with moderate deviations from solar metallicity and C/O ratio, our conclusions may not apply to planets with C/O ratios $>$1. In that case, the composition changes drastically and the abundances of CH$_4$ and HCN can become comparable to CO for both equilibrium chemistry and disequilibrium chemistry even on the day sides of planets with temperatures comparable to HD 189733b or hotter \citep{MosesEtAl2013COratio}. However, up to date no planet atmosphere has been shown to have a C/O ratio$>1$ \citep{LineEtAl2014,Benneke2015}.

We further note that our method of assuming a constant CH$_4$/CO ratio is only a good approximation for situations in which CH$_4$ and CO abundances have been homogenized horizontally and vertically throughout most of the photosphere (at pressures between $\sim$1 mbar to 1 bar). While for hot Jupiters with relatively cool equilibrium temperatures such as HD 189733b this is expected to be the case \citep{CooperShowman2006,AgundezEtAl2014,DrummondEtAl2018}, there can be situations in which abundances are homogenized only horizontally but not vertically or in which the quenching happens only at lower pressures and significant parts of the photoshpere are still in chemical equilibrium \citep{AgundezEtAl2014,MendoncaEtAl2018}. In general, such situations are expected for planets with hotter day side temperatures and weak vertical mixing. Future work is necessary to better understand when such situations occur and to clarify the relative importance of horizontal and vertical quenching.

\subsection{Other factors impacting the predicted phase curves}
\label{subsec:discussion_otherfactors}
As discussed in Section \ref{sec:phasecurvepredictions}, disequilibrium carbon chemistry barely affects the 4.5 $\mu$m phase curve while significantly decreasing the phase curve amplitude in the 3.6 $\mu$m band, thus worsening the fit of the 3.6 $\mu$m phase curve to observations. Among the simulations we ran for this paper, the equilibrium chemistry simulation actually matches the data better than any of the disequilibrium chemistry simulations. 
At this point, we would like to note that the equilibrium chemistry phase curve in this paper somewhat differs from the solar metallicity (equlibrium chemistry) phase curve presented in \citet{ShowmanEtAl2009} and \citet{KnutsonEtAl2012}. This is due to updates of the opacities \citep{FreedmanEtAl2014} and planetary and stellar parameters. In general, in all wavelength bands, the phase curve amplitude increases and the phase curve offset decreases slightly with the updated opacities. With these changes, the issue of the model overpredicting the night side fluxes in the 4.5 $\mu$m band is somewhat mitigated and difference in phase curve offset becomes the more dominant discrepancy between observations and the solar metallicity equilibrium chemistry model.

Our finding that the phase curve from the equilibrium chemistry simulation matches the observational phase curve better than the ones assuming disequilibrium chemistry is in agreement with \citet{Dobbs-DixonCowan2017} who find that based on their GCM the phase curve data are consistent with equilibrium chemistry. Does this mean that the atmosphere of HD 189733b is in chemical equilibrium? Based on the theoretical understanding of chemical kinetics and atmospheric dynamics, this is highly unlikely. It is much more likely that other processes that our model does not take into account are responsible for the shape of the phase curve. A likely explanation is the presence of clouds on the night side of HD 189733b. 
Cloud microphysics models \citep{LeeEtAl2015,PowellEtAl2018} 
and GCM simulations including clouds with varying levels of complexity \citep[e.g.][]{ParmentierEtAl2016,OreshenkoEtAl2016,LeeEtAl2016,LinesEtAl2018} 
 show that in the atmospheres of hot Jupiters there exist a variety of species that may condense to form clouds on the night side. An optically thick cloud deck on the night side would block the emission from hotter, deeper layers in the atmosphere, thus reducing the flux emitted on the night side over a broad range of wavelengths. Clouds present only on the night side of the planet would therefore increase the amplitude of the phase curve. \citet{MendoncaEtAl2018} include a simple parametrization of night side clouds in their GCM simulations of WASP-43b and study the effect on the spectra and phase curves. They are able to match the spectrally resolved HST WFC3 phase curve from 1.1-1.7 $\mu$m and the Spitzer 3.6 $\mu$m phase curve quite naturally. This demonstrates both the importance of including the effect of night side clouds and the potential of clouds to explain the low observed night side fluxes on several hot Jupiters. However, the effect of clouds on spectra depends on many unknown properties such as cloud top pressure, composition and latitudinal and longitudinal distribution and more theoretical and observational work is necessary.
Furthermore, to match the 4.5 $\mu$m phase curve, \citet{MendoncaEtAl2018} include additional CO$_2$ compared to the equilibrium chemistry abundance on the night side. There is no clear theoretical motivation for additional CO$_2$ on the night side---in contrast, \cite{AgundezEtAl2014} find that disequilibrium chemistry tends to reduce the CO$_2$ abundance on the night sides of HD 209458b and HD 189733b compared to equilibrium chemistry. \citet{MendoncaEtAl2018_chemistry} find that disequilibrium chemistry reduces the CO$_2$ abundance on the night side in the C/O=0.5 case but increases it in the C/O=2 case. They conclude that the changes in CO$_2$ due to disequilibrium chemistry cannot resolve the remaining discrepancy in the 4.5 $\mu$m band.

Especially on relatively cool planets such as HD 189733b, clouds need not be restricted to the night side. An optically thick cloud deck extending into parts of the day side or even covering the entire planet is another plausible possibility \citep[e.g.,][]{ParmentierEtAl2016,RomanRauscher2019}. Using a simple, constant particle size cloud scheme that includes radiative feedback, \citet{RomanRauscher2019} find that their cloud distribution (covering large fractions of the day side) results in a larger phase amplitude and lower phase offset than in the cloud-free case. They stress the role of cloud radiative feedback in shaping the phase curve. However, it is likely that a realistic distribution of clouds over the planets is not uniform in particle size and density \citep{LeeEtAl2016,LinesEtAl2018}, potentially resulting in a much more complex effect on the phase curve \citep{LinesEtAl2018}.

In addition to clouds, several other parameters not explored in this study can significantly impact the phase curves, including non-solar metallicities and C/O ratios and atmospheric drag due to subgrid-scale turbulence or due to Lorentz forces in a partially ionized atmosphere. Independent of these factors, there are also moderate uncertainties associated with the numerical model, most notably the uncertainty in the opacities of some species which are not well-known at high temperatures. In this context, we would like to remind the reader that the simulations presented are pure forward models and we make no attempt to fit the observational data. The general trends observed in this paper thus are much more important and meaningful than specific predictions for the emitted flux or other observable quantities.

\section{Conclusion}
We have included the radiative effect of transport-induced disequilibrium CH$_4$, CO and H$_2$O abundances in a GCM to study the effect on the atmospheric structure, phase curves and emission spectra. We have assumed that the ratio of CH$_4$ to CO is constant throughout the entire simulation (an assumption that is expected to be well fulfilled at pressures between $\sim10^{-4}$ bars and 1 bar) and treat the CH$_4$/CO ratio as free parameter. The water abundance is updated accordingly, such that the total number of oxygen atoms is preserved. It is important to include this change in the water abundance, as in equilibrium chemistry the water abundance varies by a factor of $\sim2$ between the CO-dominated day side and the CH$_4$-dominated night side. Assuming vertical and horizontal quenching, however, the water abundance is expected to be homogenized between day and night side. We ran simulations of hot Jupiter HD 189733b with eight different quenched CH$_4$/CO ratios.

We find that in the CO dominated case, which is the case favored by chemical kinetics models, the temperature changes locally by up to 150 K, with cooler temperatures compared to equilibrium chemistry on the day side and warmer temperatures on part of the night side. In the less plausible CH$_4$ dominated case, the addition of greenhouse gases leads to hotter temperatures everywhere at pressures higher than a few tens of mbars.
When comparing the predicted phase curves from GCM simulations including disequilibrium CH$_4$ and CO abundances to phase curves obtained from an equilibrium chemistry GCM simulation that has been post-processed assuming quenched CH$_4$/CO abundances, we find that the eastward offset of the phase curve maximum can differ by up to $10^\circ$.

We thus conclude that it is important to self-consistently include the effect of disequilibrium abundances of CH$_4$ and CO on the opacities in GCMs rather than including disequilibrium abundances only in the post-processing while continuing to use opacities based on equilibrium chemistry abundances in the GCM. This is in contrast to \citet{DrummondEtAl2018} who find in their study of HD 209458b that the effect of radiative feedback of disequilibrium abundances on the temperature and wind fields is only $\sim 1 \%$, but agrees with their more recent findings for HD 189733b \citep{DrummondEtAl2018HD189733b}. These seemingly conflicting results can be understood when considering the difference in the equilibrium temperatures of the planets: On the hotter HD 209458b, the CH$_4$ abundance remains low compared to the CO and H$_2$O abundances in both equilibrium and disequilibrium chemistry even on the night side. In contrast, on the cooler HD 189733b, the night side is cool enough to be dominated by CH$_4$ in equilibrium chemistry. Including disequilibrium chemistry thus changes the dominant carbon species on half of the planet, resulting in much larger changes. In addition, in the regions where disequilibrium chemistry changes the dominant carbon species, the water abundance is also altered by a factor of $\sim2$, further contributing to the effect on temperatures. 

Furthermore, we show that disequilibrium CH$_4$ and CO abundances have only a small effect on the Spitzer 4.5 $\mu$m phase curve despite CO having a prominent absorption band within this wavelength band. This is because the change in opacity due to CO is offset by a change in water opacity in the opposite direction.
In wavelength regions dominated by CH$_4$ opacity, including the Spitzer 3.6 $\mu$m and 8 $\mu$m bands, the phase curve amplitude decreases significantly, resulting in a much worse fit to the observed Spitzer 3.6 $\mu$m and 8 $\mu$m phase curves.
We thus conclude that disequilibrium carbon chemistry cannot explain the observed low night side fluxes in the 4.5 $\mu$m band, in contrast to the interpretation of \citet{KnutsonEtAl2012}. Other effects, for example night side clouds, must be responsible for the observed shape of the phase curve.

While disequilibrium chemistry does not explain existing observations of HD 189733b, it may be detectable on other hot Jupiters with a similar equilibrium temperature. Therefore, we examine the effect of disequilibrium carbon chemistry on emission spectra and simulated JWST observations. 
We find that in the expected CO dominated regime, spectral regions dominated by methane absorption bands are most suitable to observe disequilibrium abundances. Assuming that the spectral signatures of disequilibrium carbon chemistry are not obscured by clouds or other effects not considered in our model, it will be possible to distinguish between different quenched ratios with JWST in both secondary eclipse and phase curve observations. 

\acknowledgements
We thank Jonathan Fortney and Thaddeus Komacek for helpful comments and Heather Knutson for sharing the observational data for Figure \ref{fig:phasecurves_with_data}. This research was supported in part by NASA Origins grant NNX12AI79G to APS and by NASA Headquarters under the NASA Earth and Space Science Fellowship Program - Grant 80NSSC18K1248. This work benefited from the Exoplanet Summer Program in the Other Worlds Laboratory (OWL) at the University of California, Santa Cruz, a program funded by the Heising-Simons Foundation.

\bibliographystyle{aasjournal}

\begin{thebibliography}{}
\expandafter\ifx\csname natexlab\endcsname\relax\def\natexlab#1{#1}\fi
\providecommand{\url}[1]{\href{#1}{#1}}
\providecommand{\dodoi}[1]{doi:~\href{http://doi.org/#1}{\nolinkurl{#1}}}
\providecommand{\doeprint}[1]{\href{http://ascl.net/#1}{\nolinkurl{http://ascl.net/#1}}}
\providecommand{\doarXiv}[1]{\href{https://arxiv.org/abs/#1}{\nolinkurl{https://arxiv.org/abs/#1}}}

\bibitem[{{Adcroft} {et~al.}(2004){Adcroft}, {Campin}, {Hill}, \&
  {Marshall}}]{AdcroftEtAl2004}
{Adcroft}, A., {Campin}, J.-M., {Hill}, C., \& {Marshall}, J. 2004, Monthly
  Weather Review, 132, 2845, \dodoi{10.1175/MWR2823.1}

\bibitem[{{Agol} {et~al.}(2010){Agol}, {Cowan}, {Knutson}, {Deming}, {Steffen},
  {Henry}, \& {Charbonneau}}]{AgolEtAl2010}
{Agol}, E., {Cowan}, N.~B., {Knutson}, H.~A., {et~al.} 2010, \apj, 721, 1861,
  \dodoi{10.1088/0004-637X/721/2/1861}

\bibitem[{{Ag{\'u}ndez} {et~al.}(2014){Ag{\'u}ndez}, {Parmentier}, {Venot},
  {Hersant}, \& {Selsis}}]{AgundezEtAl2014}
{Ag{\'u}ndez}, M., {Parmentier}, V., {Venot}, O., {Hersant}, F., \& {Selsis},
  F. 2014, \aap, 564, A73, \dodoi{10.1051/0004-6361/201322895}

\bibitem[{{Ag{\'u}ndez} {et~al.}(2012){Ag{\'u}ndez}, {Venot}, {Iro}, {Selsis},
  {Hersant}, {H{\'e}brard}, \& {Dobrijevic}}]{AgundezEtAl2012}
{Ag{\'u}ndez}, M., {Venot}, O., {Iro}, N., {et~al.} 2012, \aap, 548, A73,
  \dodoi{10.1051/0004-6361/201220365}

\bibitem[{{Amundsen} {et~al.}(2017){Amundsen}, {Tremblin}, {Manners},
  {Baraffe}, \& {Mayne}}]{AmundsenEtAl2017}
{Amundsen}, D.~S., {Tremblin}, P., {Manners}, J., {Baraffe}, I., \& {Mayne},
  N.~J. 2017, \aap, 598, A97, \dodoi{10.1051/0004-6361/201629322}

\bibitem[{{Amundsen} {et~al.}(2016){Amundsen}, {Mayne}, {Baraffe}, {Manners},
  {Tremblin}, {Drummond}, {Smith}, {Acreman}, \& {Homeier}}]{AmundsenEtAl2016}
{Amundsen}, D.~S., {Mayne}, N.~J., {Baraffe}, I., {et~al.} 2016, \aap, 595,
  A36, \dodoi{10.1051/0004-6361/201629183}

\bibitem[{{Batalha} {et~al.}(2017){Batalha}, {Mandell}, {Pontoppidan},
  {Stevenson}, {Lewis}, {Kalirai}, {Earl}, {Greene}, {Albert}, \&
  {Nielsen}}]{BatalhaEtAl2017}
{Batalha}, N.~E., {Mandell}, A., {Pontoppidan}, K., {et~al.} 2017, \pasp, 129,
  064501, \dodoi{10.1088/1538-3873/aa65b0}

\bibitem[{{Benneke}(2015)}]{Benneke2015}
{Benneke}, B. 2015, ArXiv e-prints, arXiv:1504.07655.
\newblock \doarXiv{1504.07655}

\bibitem[{{Boyajian} {et~al.}(2015){Boyajian}, {von Braun}, {Feiden}, {Huber},
  {Basu}, {Demarque}, {Fischer}, {Schaefer}, {Mann}, {White}, {Maestro},
  {Brewer}, {Lamell}, {Spada}, {L{\'o}pez-Morales}, {Ireland}, {Farrington},
  {van Belle}, {Kane}, {Jones}, {ten Brummelaar}, {Ciardi}, {McAlister},
  {Ridgway}, {Goldfinger}, {Turner}, \& {Sturmann}}]{BoyajianEtAl2015}
{Boyajian}, T., {von Braun}, K., {Feiden}, G.~A., {et~al.} 2015, \mnras, 447,
  846, \dodoi{10.1093/mnras/stu2502}

\bibitem[{{Burrows} {et~al.}(2003){Burrows}, {Sudarsky}, \&
  {Hubbard}}]{BurrowsEtAl2003}
{Burrows}, A., {Sudarsky}, D., \& {Hubbard}, W.~B. 2003, \apj, 594, 545,
  \dodoi{10.1086/376897}

\bibitem[{{Burrows} {et~al.}(1997){Burrows}, {Marley}, {Hubbard}, {Lunine},
  {Guillot}, {Saumon}, {Freedman}, {Sudarsky}, \& {Sharp}}]{BurrowsEtAl1997}
{Burrows}, A., {Marley}, M., {Hubbard}, W.~B., {et~al.} 1997, \apj, 491, 856,
  \dodoi{10.1086/305002}

\bibitem[{{Cooper} \& {Showman}(2006)}]{CooperShowman2006}
{Cooper}, C.~S., \& {Showman}, A.~P. 2006, \apj, 649, 1048,
  \dodoi{10.1086/506312}

\bibitem[{{Dobbs-Dixon} \& {Agol}(2013)}]{Dobbs-DixonAgol2013}
{Dobbs-Dixon}, I., \& {Agol}, E. 2013, \mnras, 435, 3159,
  \dodoi{10.1093/mnras/stt1509}

\bibitem[{{Dobbs-Dixon} \& {Cowan}(2017)}]{Dobbs-DixonCowan2017}
{Dobbs-Dixon}, I., \& {Cowan}, N.~B. 2017, \apjl, 851, L26,
  \dodoi{10.3847/2041-8213/aa9bec}

\bibitem[{{Dobbs-Dixon} \& {Lin}(2008)}]{Dobbs-DixonLin2008}
{Dobbs-Dixon}, I., \& {Lin}, D.~N.~C. 2008, \apj, 673, 513,
  \dodoi{10.1086/523786}

\bibitem[{{Drummond} {et~al.}(2018{\natexlab{a}}){Drummond}, {Mayne},
  {Manners}, {Baraffe}, {Goyal}, {Tremblin}, {Sing}, \&
  {Kohary}}]{DrummondEtAl2018HD189733b}
{Drummond}, B., {Mayne}, N.~J., {Manners}, J., {et~al.} 2018{\natexlab{a}},
  \apj, 869, 28, \dodoi{10.3847/1538-4357/aaeb28}

\bibitem[{{Drummond} {et~al.}(2016){Drummond}, {Tremblin}, {Baraffe},
  {Amundsen}, {Mayne}, {Venot}, \& {Goyal}}]{DrummondEtAl2016}
{Drummond}, B., {Tremblin}, P., {Baraffe}, I., {et~al.} 2016, \aap, 594, A69,
  \dodoi{10.1051/0004-6361/201628799}

\bibitem[{{Drummond} {et~al.}(2018{\natexlab{b}}){Drummond}, {Mayne},
  {Manners}, {Carter}, {Boutle}, {Baraffe}, {H{\'e}brard}, {Tremblin}, {Sing},
  {Amundsen}, \& {Acreman}}]{DrummondEtAl2018}
{Drummond}, B., {Mayne}, N.~J., {Manners}, J., {et~al.} 2018{\natexlab{b}},
  \apj, 855, L31, \dodoi{10.3847/2041-8213/aab209}

\bibitem[{{Fortney} {et~al.}(2006){Fortney}, {Cooper}, {Showman}, {Marley}, \&
  {Freedman}}]{FortneyEtAl2006}
{Fortney}, J.~J., {Cooper}, C.~S., {Showman}, A.~P., {Marley}, M.~S., \&
  {Freedman}, R.~S. 2006, \apj, 652, 746, \dodoi{10.1086/508442}

\bibitem[{{Fortney} {et~al.}(2008){Fortney}, {Lodders}, {Marley}, \&
  {Freedman}}]{FortneyEtAl2008}
{Fortney}, J.~J., {Lodders}, K., {Marley}, M.~S., \& {Freedman}, R.~S. 2008,
  \apj, 678, 1419, \dodoi{10.1086/528370}

\bibitem[{{Fortney} {et~al.}(2005){Fortney}, {Marley}, {Lodders}, {Saumon}, \&
  {Freedman}}]{FortneyEtAl2005}
{Fortney}, J.~J., {Marley}, M.~S., {Lodders}, K., {Saumon}, D., \& {Freedman},
  R. 2005, \apjl, 627, L69, \dodoi{10.1086/431952}

\bibitem[{{Freedman} {et~al.}(2014){Freedman}, {Lustig-Yaeger}, {Fortney},
  {Lupu}, {Marley}, \& {Lodders}}]{FreedmanEtAl2014}
{Freedman}, R.~S., {Lustig-Yaeger}, J., {Fortney}, J.~J., {et~al.} 2014, \apjs,
  214, 25, \dodoi{10.1088/0067-0049/214/2/25}

\bibitem[{{Freedman} {et~al.}(2008){Freedman}, {Marley}, \&
  {Lodders}}]{FreedmanEtAl2008}
{Freedman}, R.~S., {Marley}, M.~S., \& {Lodders}, K. 2008, \apjs, 174, 504,
  \dodoi{10.1086/521793}

\bibitem[{{Goody} \& {Yung}(1989)}]{GoodyYung1989}
{Goody}, R.~M., \& {Yung}, Y.~L. 1989, {Atmospheric Radiation : Theoretical
  Basis} (Oxford University Press)

\bibitem[{{Guillot} \& {Showman}(2002)}]{GuillotShowman2002}
{Guillot}, T., \& {Showman}, A.~P. 2002, \aap, 385, 156,
  \dodoi{10.1051/0004-6361:20011624}

\bibitem[{{Hauschildt} {et~al.}(1999){Hauschildt}, {Allard}, \&
  {Baron}}]{HauschildtEtAl1999}
{Hauschildt}, P.~H., {Allard}, F., \& {Baron}, E. 1999, \apj, 512, 377,
  \dodoi{10.1086/306745}

\bibitem[{{Heng} {et~al.}(2011){Heng}, {Frierson}, \&
  {Phillipps}}]{HengEtAl2011b}
{Heng}, K., {Frierson}, D. M.~W., \& {Phillipps}, P.~J. 2011, \mnras, 418,
  2669, \dodoi{10.1111/j.1365-2966.2011.19658.x}

\bibitem[{{Kataria} {et~al.}(2014){Kataria}, {Showman}, {Fortney}, {Marley}, \&
  {Freedman}}]{KatariaEtAl2014}
{Kataria}, T., {Showman}, A.~P., {Fortney}, J.~J., {Marley}, M.~S., \&
  {Freedman}, R.~S. 2014, \apj, 785, 92, \dodoi{10.1088/0004-637X/785/2/92}

\bibitem[{{Kataria} {et~al.}(2015){Kataria}, {Showman}, {Fortney}, {Stevenson},
  {Line}, {Kreidberg}, {Bean}, \& {D{\'e}sert}}]{KatariaEtAl2015}
{Kataria}, T., {Showman}, A.~P., {Fortney}, J.~J., {et~al.} 2015, \apj, 801,
  86, \dodoi{10.1088/0004-637X/801/2/86}

\bibitem[{{Kataria} {et~al.}(2013){Kataria}, {Showman}, {Lewis}, {Fortney},
  {Marley}, \& {Freedman}}]{KatariaEtAl2013}
{Kataria}, T., {Showman}, A.~P., {Lewis}, N.~K., {et~al.} 2013, \apj, 767, 76,
  \dodoi{10.1088/0004-637X/767/1/76}

\bibitem[{{Kataria} {et~al.}(2016){Kataria}, {Sing}, {Lewis}, {Visscher},
  {Showman}, {Fortney}, \& {Marley}}]{KatariaEtAl2016}
{Kataria}, T., {Sing}, D.~K., {Lewis}, N.~K., {et~al.} 2016, \apj, 821, 9,
  \dodoi{10.3847/0004-637X/821/1/9}

\bibitem[{{Knutson} {et~al.}(2007){Knutson}, {Charbonneau}, {Allen}, {Fortney},
  {Agol}, {Cowan}, {Showman}, {Cooper}, \& {Megeath}}]{KnutsonEtAl2007}
{Knutson}, H.~A., {Charbonneau}, D., {Allen}, L.~E., {et~al.} 2007, \nat, 447,
  183, \dodoi{10.1038/nature05782}

\bibitem[{{Knutson} {et~al.}(2009){Knutson}, {Charbonneau}, {Cowan}, {Fortney},
  {Showman}, {Agol}, {Henry}, {Everett}, \& {Allen}}]{KnutsonEtAl2009_HD189}
{Knutson}, H.~A., {Charbonneau}, D., {Cowan}, N.~B., {et~al.} 2009, \apj, 690,
  822, \dodoi{10.1088/0004-637X/690/1/822}

\bibitem[{{Knutson} {et~al.}(2012){Knutson}, {Lewis}, {Fortney}, {Burrows},
  {Showman}, {Cowan}, {Agol}, {Aigrain}, {Charbonneau}, {Deming}, {D{\'e}sert},
  {Henry}, {Langton}, \& {Laughlin}}]{KnutsonEtAl2012}
{Knutson}, H.~A., {Lewis}, N., {Fortney}, J.~J., {et~al.} 2012, \apj, 754, 22,
  \dodoi{10.1088/0004-637X/754/1/22}

\bibitem[{{Komacek} \& {Youdin}(2017)}]{KomacekYoudin2017}
{Komacek}, T.~D., \& {Youdin}, A.~N. 2017, \apj, 844, 94,
  \dodoi{10.3847/1538-4357/aa7b75}

\bibitem[{{Kreidberg} {et~al.}(2014){Kreidberg}, {Bean}, {D{\'e}sert},
  {Benneke}, {Deming}, {Stevenson}, {Seager}, {Berta-Thompson}, {Seifahrt}, \&
  {Homeier}}]{KreidbergEtAl2014}
{Kreidberg}, L., {Bean}, J.~L., {D{\'e}sert}, J.-M., {et~al.} 2014, \nat, 505,
  69, \dodoi{10.1038/nature12888}

\bibitem[{{Lee} {et~al.}(2016){Lee}, {Dobbs-Dixon}, {Helling}, {Bognar}, \&
  {Woitke}}]{LeeEtAl2016}
{Lee}, G., {Dobbs-Dixon}, I., {Helling}, C., {Bognar}, K., \& {Woitke}, P.
  2016, \aap, 594, A48, \dodoi{10.1051/0004-6361/201628606}

\bibitem[{{Lee} {et~al.}(2015){Lee}, {Helling}, {Dobbs-Dixon}, \&
  {Juncher}}]{LeeEtAl2015}
{Lee}, G., {Helling}, C., {Dobbs-Dixon}, I., \& {Juncher}, D. 2015, \aap, 580,
  A12, \dodoi{10.1051/0004-6361/201525982}

\bibitem[{{Lewis} {et~al.}(2014){Lewis}, {Showman}, {Fortney}, {Knutson}, \&
  {Marley}}]{LewisEtAl2014}
{Lewis}, N.~K., {Showman}, A.~P., {Fortney}, J.~J., {Knutson}, H.~A., \&
  {Marley}, M.~S. 2014, \apj, 795, 150, \dodoi{10.1088/0004-637X/795/2/150}

\bibitem[{{Lewis} {et~al.}(2010){Lewis}, {Showman}, {Fortney}, {Marley},
  {Freedman}, \& {Lodders}}]{LewisEtAl2010}
{Lewis}, N.~K., {Showman}, A.~P., {Fortney}, J.~J., {et~al.} 2010, \apj, 720,
  344, \dodoi{10.1088/0004-637X/720/1/344}

\bibitem[{{Line} {et~al.}(2014){Line}, {Knutson}, {Wolf}, \&
  {Yung}}]{LineEtAl2014}
{Line}, M.~R., {Knutson}, H., {Wolf}, A.~S., \& {Yung}, Y.~L. 2014, \apj, 783,
  70, \dodoi{10.1088/0004-637X/783/2/70}

\bibitem[{{Lines} {et~al.}(2018){Lines}, {Mayne}, {Boutle}, {Manners}, {Lee},
  {Helling}, {Drummond}, {Amundsen}, {Goyal}, {Acreman}, {Tremblin}, \&
  {Kerslake}}]{LinesEtAl2018}
{Lines}, S., {Mayne}, N.~J., {Boutle}, I.~A., {et~al.} 2018, \aap, 615, A97,
  \dodoi{10.1051/0004-6361/201732278}

\bibitem[{{Lodders} \& {Fegley}(2002)}]{LoddersFegley2002}
{Lodders}, K., \& {Fegley}, B. 2002, \icarus, 155, 393,
  \dodoi{10.1006/icar.2001.6740}

\bibitem[{{Marley} \& {McKay}(1999)}]{MarleyMcKay1999}
{Marley}, M.~S., \& {McKay}, C.~P. 1999, \icarus, 138, 268,
  \dodoi{10.1006/icar.1998.6071}

\bibitem[{{Marley} {et~al.}(1996){Marley}, {Saumon}, {Guillot}, {Freedman},
  {Hubbard}, {Burrows}, \& {Lunine}}]{MarleyEtAl1996}
{Marley}, M.~S., {Saumon}, D., {Guillot}, T., {et~al.} 1996, Science, 272,
  1919, \dodoi{10.1126/science.272.5270.1919}

\bibitem[{{Marley} {et~al.}(2002){Marley}, {Seager}, {Saumon}, {Lodders},
  {Ackerman}, {Freedman}, \& {Fan}}]{MarleyEtAl2002}
{Marley}, M.~S., {Seager}, S., {Saumon}, D., {et~al.} 2002, \apj, 568, 335,
  \dodoi{10.1086/338800}

\bibitem[{{Mayne} {et~al.}(2014){Mayne}, {Baraffe}, {Acreman}, {Smith},
  {Browning}, {Sk{\r{a}}lid Amundsen}, {Wood}, {Thuburn}, \&
  {Jackson}}]{MayneEtAl2014}
{Mayne}, N.~J., {Baraffe}, I., {Acreman}, D.~M., {et~al.} 2014, \aap, 561,
  \dodoi{10.1051/0004-6361/201322174}

\bibitem[{{McKay} {et~al.}(1989){McKay}, {Pollack}, \&
  {Courtin}}]{McKayEtAl1989}
{McKay}, C.~P., {Pollack}, J.~B., \& {Courtin}, R. 1989, \icarus, 80, 23,
  \dodoi{10.1016/0019-1035(89)90160-7}

\bibitem[{{Mendon{\c{c}}a} {et~al.}(2016){Mendon{\c{c}}a}, {Grimm},
  {Grosheintz}, \& {Heng}}]{MendoncaEtAl2016}
{Mendon{\c{c}}a}, J.~M., {Grimm}, S.~L., {Grosheintz}, L., \& {Heng}, K. 2016,
  \apj, 829, 115, \dodoi{10.3847/0004-637X/829/2/115}

\bibitem[{{Mendon{\c{c}}a} {et~al.}(2018{\natexlab{a}}){Mendon{\c{c}}a},
  {Malik}, {Demory}, \& {Heng}}]{MendoncaEtAl2018}
{Mendon{\c{c}}a}, J.~M., {Malik}, M., {Demory}, B.-O., \& {Heng}, K.
  2018{\natexlab{a}}, \aj, 155, 150, \dodoi{10.3847/1538-3881/aaaebc}

\bibitem[{{Mendon{\c{c}}a} {et~al.}(2018{\natexlab{b}}){Mendon{\c{c}}a},
  {Tsai}, {Malik}, {Grimm}, \& {Heng}}]{MendoncaEtAl2018_chemistry}
{Mendon{\c{c}}a}, J.~M., {Tsai}, S.-m., {Malik}, M., {Grimm}, S.~L., \& {Heng},
  K. 2018{\natexlab{b}}, \apj, 869, 107, \dodoi{10.3847/1538-4357/aaed23}

\bibitem[{{Menou}(2019)}]{Menou2019}
{Menou}, K. 2019, \mnras, 485, L98, \dodoi{10.1093/mnrasl/slz041}

\bibitem[{{Miguel} \& {Kaltenegger}(2014)}]{MiguelKaltenegger2014}
{Miguel}, Y., \& {Kaltenegger}, L. 2014, \apj, 780, 166,
  \dodoi{10.1088/0004-637X/780/2/166}

\bibitem[{{Moses} {et~al.}(2013{\natexlab{a}}){Moses}, {Madhusudhan},
  {Visscher}, \& {Freedman}}]{MosesEtAl2013COratio}
{Moses}, J.~I., {Madhusudhan}, N., {Visscher}, C., \& {Freedman}, R.~S.
  2013{\natexlab{a}}, \apj, 763, 25, \dodoi{10.1088/0004-637X/763/1/25}

\bibitem[{{Moses} {et~al.}(2011){Moses}, {Visscher}, {Fortney}, {Showman},
  {Lewis}, {Griffith}, {Klippenstein}, {Shabram}, {Friedson}, {Marley}, \&
  {Freedman}}]{MosesEtAl2011}
{Moses}, J.~I., {Visscher}, C., {Fortney}, J.~J., {et~al.} 2011, \apj, 737, 15,
  \dodoi{10.1088/0004-637X/737/1/15}

\bibitem[{{Moses} {et~al.}(2013{\natexlab{b}}){Moses}, {Line}, {Visscher},
  {Richardson}, {Nettelmann}, {Fortney}, {Barman}, {Stevenson}, \&
  {Madhusudhan}}]{MosesEtAl2013GJ436b}
{Moses}, J.~I., {Line}, M.~R., {Visscher}, C., {et~al.} 2013{\natexlab{b}},
  \apj, 777, 34, \dodoi{10.1088/0004-637X/777/1/34}

\bibitem[{{Oreshenko} {et~al.}(2016){Oreshenko}, {Heng}, \&
  {Demory}}]{OreshenkoEtAl2016}
{Oreshenko}, M., {Heng}, K., \& {Demory}, B.-O. 2016, \mnras, 457, 3420,
  \dodoi{10.1093/mnras/stw133}

\bibitem[{{Parmentier} {et~al.}(2016){Parmentier}, {Fortney}, {Showman},
  {Morley}, \& {Marley}}]{ParmentierEtAl2016}
{Parmentier}, V., {Fortney}, J.~J., {Showman}, A.~P., {Morley}, C., \&
  {Marley}, M.~S. 2016, \apj, 828, 22, \dodoi{10.3847/0004-637X/828/1/22}

\bibitem[{{Parmentier} {et~al.}(2013){Parmentier}, {Showman}, \&
  {Lian}}]{ParmentierEtAl2013}
{Parmentier}, V., {Showman}, A.~P., \& {Lian}, Y. 2013, \aap, 558, A91,
  \dodoi{10.1051/0004-6361/201321132}

\bibitem[{{Parmentier} {et~al.}(2018){Parmentier}, {Line}, {Bean}, {Mansfield},
  {Kreidberg}, {Lupu}, {Visscher}, {D{\'e}sert}, {Fortney}, {Deleuil},
  {Arcangeli}, {Showman}, \& {Marley}}]{ParmentierEtAl2018}
{Parmentier}, V., {Line}, M.~R., {Bean}, J.~L., {et~al.} 2018, \aap, 617, A110,
  \dodoi{10.1051/0004-6361/201833059}

\bibitem[{{Perna} {et~al.}(2012){Perna}, {Heng}, \& {Pont}}]{PernaEtAl2012}
{Perna}, R., {Heng}, K., \& {Pont}, F. 2012, \apj, 751, 59,
  \dodoi{10.1088/0004-637X/751/1/59}

\bibitem[{{Pontoppidan} {et~al.}(2016){Pontoppidan}, {Pickering}, {Laidler},
  {Gilbert}, {Sontag}, {Slocum}, {Sienkiewicz}, {Hanley}, {Earl}, {Pueyo},
  {Ravindranath}, {Karakla}, {Robberto}, {Noriega-Crespo}, \&
  {Barker}}]{PontoppidanEtAl2016}
{Pontoppidan}, K.~M., {Pickering}, T.~E., {Laidler}, V.~G., {et~al.} 2016, in
  \procspie, Vol. 9910, Observatory Operations: Strategies, Processes, and
  Systems VI, 991016

\bibitem[{{Powell} {et~al.}(2018){Powell}, {Zhang}, {Gao}, \&
  {Parmentier}}]{PowellEtAl2018}
{Powell}, D., {Zhang}, X., {Gao}, P., \& {Parmentier}, V. 2018, \apj, 860, 18,
  \dodoi{10.3847/1538-4357/aac215}

\bibitem[{{Rauscher} \& {Menou}(2012)}]{RauscherMenou2012}
{Rauscher}, E., \& {Menou}, K. 2012, \apj, 750, 96,
  \dodoi{10.1088/0004-637X/750/2/96}

\bibitem[{{Roman} \& {Rauscher}(2019)}]{RomanRauscher2019}
{Roman}, M., \& {Rauscher}, E. 2019, \apj, 872, 1,
  \dodoi{10.3847/1538-4357/aafdb5}

\bibitem[{{Shapiro}(1970)}]{Shapiro1970}
{Shapiro}, R. 1970, Reviews of Geophysics and Space Physics, 8, 359,
  \dodoi{10.1029/RG008i002p00359}

\bibitem[{{Sharp} \& {Burrows}(2007)}]{SharpBurrows2007}
{Sharp}, C.~M., \& {Burrows}, A. 2007, \apjs, 168, 140, \dodoi{10.1086/508708}

\bibitem[{{Showman} {et~al.}(2013){Showman}, {Fortney}, {Lewis}, \&
  {Shabram}}]{ShowmanEtAl2013}
{Showman}, A.~P., {Fortney}, J.~J., {Lewis}, N.~K., \& {Shabram}, M. 2013,
  \apj, 762, 24, \dodoi{10.1088/0004-637X/762/1/24}

\bibitem[{{Showman} {et~al.}(2009){Showman}, {Fortney}, {Lian}, {Marley},
  {Freedman}, {Knutson}, \& {Charbonneau}}]{ShowmanEtAl2009}
{Showman}, A.~P., {Fortney}, J.~J., {Lian}, Y., {et~al.} 2009, \apj, 699, 564,
  \dodoi{10.1088/0004-637X/699/1/564}

\bibitem[{{Showman} \& {Guillot}(2002)}]{ShowmanGuillot2002}
{Showman}, A.~P., \& {Guillot}, T. 2002, \aap, 385, 166,
  \dodoi{10.1051/0004-6361:20020101}

\bibitem[{{Showman} {et~al.}(2015){Showman}, {Lewis}, \&
  {Fortney}}]{ShowmanEtAl2015}
{Showman}, A.~P., {Lewis}, N.~K., \& {Fortney}, J.~J. 2015, \apj, 801, 95,
  \dodoi{10.1088/0004-637X/801/2/95}

\bibitem[{{Showman} \& {Polvani}(2011)}]{ShowmanPolvani2011}
{Showman}, A.~P., \& {Polvani}, L.~M. 2011, \apj, 738,
  \dodoi{10.1088/0004-637X/738/1/71}

\bibitem[{{Stevenson} {et~al.}(2017){Stevenson}, {Line}, {Bean}, {D{\'e}sert},
  {Fortney}, {Showman}, {Kataria}, {Kreidberg}, \& {Feng}}]{StevensonEtAl2017}
{Stevenson}, K.~B., {Line}, M.~R., {Bean}, J.~L., {et~al.} 2017, \aj, 153, 68,
  \dodoi{10.3847/1538-3881/153/2/68}

\bibitem[{{Thrastarson} \& {Cho}(2011)}]{ThrastarsonCho2011}
{Thrastarson}, H.~T., \& {Cho}, J. Y.-K. 2011, \apj, 729, 117,
  \dodoi{10.1088/0004-637X/729/2/117}

\bibitem[{{Toon} {et~al.}(1989){Toon}, {McKay}, {Ackerman}, \&
  {Santhanam}}]{ToonEtAl1989}
{Toon}, O.~B., {McKay}, C.~P., {Ackerman}, T.~P., \& {Santhanam}, K. 1989,
  \jgr, 94, 16287, \dodoi{10.1029/JD094iD13p16287}

\bibitem[{{Tremblin} {et~al.}(2017){Tremblin}, {Chabrier}, {Mayne}, {Amundsen},
  {Baraffe}, {Debras}, {Drummond}, {Manners}, \& {Fromang}}]{TremblinEtAl2017}
{Tremblin}, P., {Chabrier}, G., {Mayne}, N.~J., {et~al.} 2017, \apj, 841, 30,
  \dodoi{10.3847/1538-4357/aa6e57}

\bibitem[{{Tsai} {et~al.}(2018){Tsai}, {Kitzmann}, {Lyons}, {Mendon{\c{c}}a},
  {Grimm}, \& {Heng}}]{TsaiEtAl2017}
{Tsai}, S.-M., {Kitzmann}, D., {Lyons}, J.~R., {et~al.} 2018, \apj, 862, 31,
  \dodoi{10.3847/1538-4357/aac834}

\bibitem[{{Venot} {et~al.}(2014){Venot}, {Ag{\'u}ndez}, {Selsis}, {Tessenyi},
  \& {Iro}}]{VenotEtAl2014}
{Venot}, O., {Ag{\'u}ndez}, M., {Selsis}, F., {Tessenyi}, M., \& {Iro}, N.
  2014, \aap, 562, A51, \dodoi{10.1051/0004-6361/201322485}

\bibitem[{{Visscher} {et~al.}(2006){Visscher}, {Lodders}, \&
  {Fegley}}]{VisscherEtAl2006}
{Visscher}, C., {Lodders}, K., \& {Fegley}, Jr., B. 2006, \apj, 648, 1181,
  \dodoi{10.1086/506245}

\bibitem[{{Visscher} \& {Moses}(2011)}]{VisscherMoses2011}
{Visscher}, C., \& {Moses}, J.~I. 2011, \apj, 738, 72,
  \dodoi{10.1088/0004-637X/738/1/72}

\bibitem[{{Zellem} {et~al.}(2014){Zellem}, {Lewis}, {Knutson}, {Griffith},
  {Showman}, {Fortney}, {Cowan}, {Agol}, {Burrows}, {Charbonneau}, {Deming},
  {Laughlin}, \& {Langton}}]{ZellemEtAl2014}
{Zellem}, R.~T., {Lewis}, N.~K., {Knutson}, H.~A., {et~al.} 2014, \apj, 790,
  53, \dodoi{10.1088/0004-637X/790/1/53}

\end{thebibliography}

\end{document}